\documentclass[aip,reprint,amsmath,amssymb,floatfix,raggedbottom,citeautoscript]{revtex4-1}
\usepackage{tikz}
\usepackage{graphicx}
\usepackage{braket}
\usepackage{txfonts}
\usepackage{bm}
\usepackage{color}
\usepackage{verbatim}
\definecolor{myblue}{rgb}{0,0,1}

\usepackage[breaklinks=true,colorlinks=true,linkcolor=myblue,urlcolor=myblue,citecolor=myblue]{hyperref}

\begin{document}

\title{Multireference perturbation theory with nonorthogonal determinants}
\author{Alex Krotz}
\author{K. Birgitta Whaley}
\author{Martin Head-Gordon}
\email{mhg@cchem.berkeley.edu}
\affiliation{Department of Chemistry, University of California, Berkeley, California 94720, United
States}

\begin{abstract}
We introduce NOCI-RSPT, a framework for constructing perturb-then-diagonalize multireference perturbation theories for model spaces defined by nonorthogonal determinants.
NOCI-RSPT realizes order-by-order size consistency by orthogonalizing the effective Hamiltonian before perturbatively truncating it and removes problematic internal excitations through model-space projection.
The resulting second- and third-order theories, NOCI-RS2 and NOCI-RS3, recover MP2 and MP3 in the single-reference limit and exhibit the same respective cost scaling at fixed model-space size.
Calculations on assemblies of multireference H$_2$ fragments demonstrate size consistency, while calculations of the low-energy states of H$_2$ along its dissociation curve demonstrate improved numerical stability and characterize the influence of intruder states.
The latter calculations demonstrate the exceptional accuracy achieved by NOCI-RS3 and illustrate how model-space expansion and amplitude regularization can mitigate the effects of large perturbative corrections.
Altogether, NOCI-RSPT provides a systematic route to low-scaling multireference perturbation theories built from single-reference methods.
\end{abstract}

\maketitle

\section{Introduction}

Reliable predictions of molecular energies and properties often require an accurate treatment of electronic correlation.
While single-reference methods often suffice, systems exhibiting both significant static and dynamic correlation generally necessitate a multireference approach or a higher-than-computationally feasible level of excitation.\cite{coeInvestigatingMultireferenceCharacter2015,lyakhMultireferenceNatureChemistry2012,khedkarModernMultireferenceMethods2021}
Multireference perturbation theories use model spaces containing multiple electronic configurations to capture the static correlation structure of the system, allowing the remaining dynamic correlation to be described perturbatively.\cite{anderssonSecondorderPerturbationTheory1990,angeliIntroductionNelectronValence2001,angeliNelectronValenceState2002,evangelistaCompanionPerturbationTheory2009,rolikPerturbationMulticonfigurationWave2003,szabadosMulticonfigurationPerturbationTheory2005}
In many approaches, the model space is defined in terms of a complete or restricted active space wherein selected electrons are distributed among a shared set of active orbitals.\cite{siegbahnCompleteActiveSpace1981,malmqvistRestrictedActiveSpace1990}
While this is a powerful strategy, the number of configurations making up the model space grows combinatorially with the size of the active space, rendering the approach prohibitive for large active spaces, even when insignificant determinants are pruned.\cite{levineCASSCFExtremelyLarge2020}
Furthermore, selecting an appropriate active space often requires system-specific expertise, especially when the character of relevant configurations changes significantly between different states or along a reaction coordinate.\cite{steinAutomatedSelectionActive2016,bensbergCorrespondingActiveOrbital2023}
Constructing both an accurate and compact model space thus remains a practical bottleneck in applying multireference theories to complex molecular systems.

Nonorthogonal determinant expansions can provide an efficient representation of multiconfigurational model spaces.
Because each determinant is expressed in its own optimized orbital basis, configuration-specific orbital relaxation is incorporated directly instead of indirectly via additional configurations.\cite{yostSizeConsistentFormulations2016}
Although determining a suitable set of nonorthogonal determinants remains a practical challenge, numerous numerical strategies have been developed to find low-energy determinants that may be good candidates, including SCF metadynamics,\cite{thomLocatingMultipleSelfConsistent2008} holomorphic Hartree--Fock theory,\cite{hiscockHolomorphichartreeFock2014a} and a wide range of orbital optimization methods.\cite{gilbertSelfConsistentFieldCalculations2008,khaliullinEfficientSelfconsistentField2006,haitExcitedStateOrbital2020}
Established and emerging techniques, including those based on nonorthogonal configuration interaction (NOCI)\cite{thomhartreeFockSolutions2009,sundstromNonorthogonalConfigurationInteraction2014,oosterbaanNonorthogonalConfigurationInteraction2018,lemkeNonorthogonalConfigurationInteraction2025,sunSelectedNonorthogonalConfiguration2024,j.mayhallSpinFlipNonorthogonal2014,sanchez-mansillaRoleDynamicElectron2022} and resonating Hartree--Fock\cite{fukutomeTheoryResonatingQuantum1988,tomitaInitioMolecularOrbital1996,niteLowCostMolecularExcited2019,niteEfficientMultiConfigurationalWavefunction,millerNumericallyStableResonating2025}, have further demonstrated the ability of nonorthogonal determinant expansions to efficiently describe correlation in diverse chemical systems.
Furthermore, recent calculations of molecular ground- and excited-state energies have demonstrated that high accuracy can be achieved with comparatively few nonorthogonal determinants.\cite{giulianiPreciseQuantumChemistry2026,giulianiVariationalLowenergySubspaces2026}
These results motivate the use of compact nonorthogonal determinant expansions as model spaces for multireference correlation theories.

Several approaches have been explored to this end.
Notably, the NOCI-PT2 method uses second-order perturbation theory to dynamically correlate a zeroth-order NOCI wavefunction.\cite{burtonReachingFullCorrelation2020}
In contrast, the $\Delta$SCF(2) and subsequent NOCI-MP2 methods take a perturb-then-diagonalize approach in which each nonorthogonal determinant is independently dressed using M\o ller--Plesset (MP) perturbation theory, after which the dressed states are coupled through an effective generalized eigenvalue problem.\cite{yostMultireferencePerturbationMethod2013,yostSizeConsistentFormulations2016}
This strategy is attractive because it allows dynamic correlation to modulate the mixing of the nonorthogonal determinants. 
Additionally, it reuses inexpensive and well-established single-reference correlation theories to obtain perturbative dressings and provides access to several correlated states through the final diagonalization.
Additionally, for a fixed model-space size, NOCI-MP2 has the same cost scaling with respect to orbital count as MP2\cite{yostSizeConsistentFormulations2016} and can have its memory overhead considerably reduced by using the resolution-of-the-identity approximation.\cite{whittenCoulombicPotentialEnergy1973,jafriElectronRepulsionIntegral1974,yostEfficientImplementationNOCIMP22018}

Despite this appeal, specifying a perturbative order in NOCI-MP2 does not uniquely determine the construction of the effective Hamiltonian and overlap matrices.
Independently truncating these matrices generally retains contributions beyond the nominal order in the corresponding orthogonalized effective Hamiltonian.
This flexibility was used to construct two variants of the original $\Delta$SCF(2) formulation that restore size consistency for a single multireference fragment in a single-reference environment.\cite{yostSizeConsistentFormulations2016}
While all three formulations of NOCI-MP2 can reasonably be described as second order, they differ substantially in their construction and, as we demonstrate below, exhibit dramatic differences in numerical stability.

Two additional difficulties arise when single-reference perturbation theories are used to construct a multireference effective Hamiltonian.
First, while a given determinant may represent an essential component of static correlation and thus warrant inclusion in the model space, it may give rise to large perturbative corrections.
Such behavior can arise due to small or even vanishing orbital-energy gaps that appear in the denominators of perturbative amplitudes, causing them to become large or singular along a dissociation curve or reaction coordinate.\cite{leeRegularizedOrbitalOptimizedSecondOrder2018}
Prior applications of NOCI-MP2 have used empirical level shifts to control this behavior, motivating a more systematic consideration of regularization within the perturb-then-diagonalize framework.\cite{yostMultireferencePerturbationMethod2013,yostSizeConsistentFormulations2016}
Second, when a model-space determinant is perturbatively correlated, its correction may incorporate excitations that lie within the span of other determinants in the model space.
These internal excitations can reproduce correlation already accounted for by mixing within the model space, introducing double counting in the effective Hamiltonian and poor conditioning among the dressed states.
In prior work, potential double counting was mitigated by omitting selected active-space excitations from the perturbative corrections, improving accuracy in selected cases.\cite{yostSizeConsistentFormulations2016}
These observations motivate a more general framework for eliminating internal excitations.
Indeed, work related to separating internal and external spaces in the context of post-NOCI correlation theories is ongoing.\cite{demoraesNaturalExcitationFramework2026}
Here, we address its specific manifestation within the perturb-then-diagonalize construction of multireference perturbation theory.

In this work, we introduce NOCI-RSPT, a framework for constructing perturb-then-diagonalize multireference perturbation theories for model spaces defined by nonorthogonal determinants.
We first formulate model-space projection for determinant-specific perturbative corrections and develop a consistent perturbative expansion of the orthogonalized effective Hamiltonian.
We establish the size consistency of this construction and use its single-reference limit to derive the second- and third-order theories that we denote as NOCI-RS2 and NOCI-RS3.
We then compare these theories with previous NOCI-MP2 formulations using assemblies of multireference H$_2$ fragments and the low-energy states of H$_2$ along its dissociation curve.
Finally, we examine the influence of large perturbative corrections and their treatment through model-space expansion and amplitude regularization.
Together, these developments provide a systematic foundation for low-scaling perturb-then-diagonalize correlation methods constructed in compact nonorthogonal model spaces.

\section{Theory}

As a starting point for formulating a multireference theory, we consider the construction of a suitable set of dressed states. 
Assuming that we have applied a single-reference Rayleigh--Schr\"{o}dinger perturbation theory (RSPT), we have access to reference-dependent waveoperators $\hat{\Omega}_{A}(\lambda)=\hat{I}+\lambda\hat{\Omega}_{A}^{(1)}+O(\lambda^2)$ through some order $n$.
Here, $\hat{\Omega}_{A}^{(k>0)}$ is composed of pure excitations with respect to the normalized reference determinant $\vert A^{(0)}\rangle$.
Acting with the waveoperator on the reference yields the unprojected dressed state $\hat{\Omega}_{A}(\lambda)\vert A^{(0)}\rangle=\vert A^{(0)}\rangle+\lambda\vert A_{\mathrm u}^{(1)}\rangle+O(\lambda^{2})$, where the subscript $\mathrm u$ denotes an unprojected correction.
In single-reference RSPT, the model space is defined by the span of the reference determinant and described by the projector $\hat{P}=\vert A^{(0)}\rangle\langle A^{(0)}\vert$.
Intermediate normalization against the reference amounts to the condition $\hat{P}\vert A(\lambda)\rangle=\vert A^{(0)}\rangle$ and ensures that the wavefunction corrections lie outside of the model space.
This notion can be generalized to a multireference model space with projector $\hat{P}=\sum_{AB}\vert A^{(0)}\rangle\big[(\mathbf{S}^{(0,0)})^{-1}\big]_{AB}\langle B^{(0)}\vert$, where, once again, we require that the dressed states satisfy $\hat{P}\vert A(\lambda)\rangle=\vert A^{(0)}\rangle$.
Notably, this condition is not generally satisfied by the unprojected dressed state because the single-reference waveoperator generates components along other model-space directions.
Our approach for constructing projected dressed states that satisfy this condition is to first construct a model-space state by applying the inverse of the model-space representation of the waveoperator to the model-space determinant, $\vert\psi_{A}(\lambda)\rangle = [\hat{P}\hat{\Omega}_{A}(\lambda)\hat{P}]_{P}^{-1}\vert A^{(0)}\rangle$, where the inverse is taken within the model space.
Then, acting with the waveoperator on $\vert\psi_{A}(\lambda)\rangle$ yields the projected dressed state
\begin{align}
    \vert A(\lambda)\rangle
    &= \hat{\Omega}_{A}(\lambda)\big[\hat{P}\hat{\Omega}_{A}(\lambda)\hat{P}\big]_{P}^{-1}\vert A^{(0)}\rangle\label{eq:ket}\\
    &=\vert A^{(0)}\rangle+\lambda (1-\hat{P})\hat{\Omega}_{A}^{(1)}\vert A^{(0)}\rangle+O(\lambda^{2}),\nonumber
\end{align}
where the zeroth-order contribution is the corresponding model-space determinant and all higher-order corrections lie outside of the model space.

Using these projected dressed states, we define the nonorthogonal effective Hamiltonian and overlap matrices with elements
\begin{align}
    S_{AB}(\mu,\nu) = \langle A(\mu)\vert B(\nu)\rangle, \label{eq:S_AB}
\end{align}
and
\begin{align}
    H_{AB}(\mu,\nu) = \langle A(\mu)\vert\hat{H}\vert B(\nu)\rangle, \label{eq:H_AB}
\end{align}
where $\mu$ and $\nu$ track the bra and ket orders independently.
Separately truncating these matrices would generally retain contributions beyond the nominal truncation order in the corresponding orthogonalized effective Hamiltonian.
We avoid this ambiguity by first transforming to an orthonormal basis, in which all perturbative corrections to the overlap matrix vanish.
A suitable orthogonalizing transformation is
\begin{align}
    \mathbf{X}(\mu,\nu) = (\mathbf{S}^{(0,0)})^{-\frac{1}{2}}
    \big[(\mathbf{S}^{(0,0)})^{-\frac{1}{2}}
    \mathbf{S}(\mu,\nu)
    (\mathbf{S}^{(0,0)})^{-\frac{1}{2}}\big]^{-\frac{1}{2}},
\end{align}
which can be expanded using the binomial series.
Applying this transformation to the overlap matrix yields $\tilde{\mathbf{S}}(\mu,\nu)=\mathbf{X}^{\dagger}(\nu,\mu)\mathbf{S}(\mu,\nu)\mathbf{X}(\mu,\nu)=\mathbf{I}$, as desired.
Similarly, the orthogonalized effective Hamiltonian is
\begin{align}
    \tilde{\mathbf{H}}(\mu,\nu)=\mathbf{X}^{\dagger}(\nu,\mu)\mathbf{H}(\mu,\nu)\mathbf{X}(\mu,\nu),\label{eq:Ht}
\end{align}
which has two-parameter coefficients denoted $\tilde{\mathbf{H}}^{(i,j)}$ that can be assembled to construct the effective multireference Hamiltonian.

Before detailing this construction, we first demonstrate that every coefficient of Eq.~\ref{eq:Ht}, and therefore any linear combination of these coefficients, is size consistent.
For noninteracting fragments with local Hilbert spaces labeled $1$ and $2$, this amounts to showing that the orthogonalized effective Hamiltonian factorizes as $\tilde{\mathbf{H}}(\mu,\nu)=\tilde{\mathbf{H}}_{1}(\mu,\nu)\otimes\mathbf{I}_{2}+\mathbf{I}_{1}\otimes\tilde{\mathbf{H}}_{2}(\mu,\nu)$.
As a necessary condition, we assume that the set of model-space determinants is product-closed such that $\{\vert A^{(0)}\rangle\}=\{\vert A_{1}^{(0)}\rangle\}\otimes\{\vert A_{2}^{(0)}\rangle\}$ and consequently that the model-space projector factorizes as $\hat{P}=\hat{P}_{1}\otimes\hat{P}_{2}$.
This is the same assumption required for the NOCI of the model-space determinants to be size consistent.\cite{sundstromNonorthogonalConfigurationInteraction2014}
As a result, the waveoperator associated with each reference determinant is multiplicatively separable, $\hat{\Omega}_{A}(\lambda) = \hat{\Omega}_{A_{1}}(\lambda)\otimes\hat{\Omega}_{A_{2}}(\lambda)$. For non-interacting (i.e., infinitely separated) fragments, the parent Hamiltonian is additively separable, $\hat{H}=\hat{H}_{1}\otimes\hat{I}_{2}+\hat{I}_{1}\otimes\hat{H}_{2}$.
Thus, the dressed states defined in Eq.~\ref{eq:ket} are multiplicatively separable, $\vert A(\lambda)\rangle=\vert A_{1}(\lambda)\rangle\otimes\vert A_{2}(\lambda)\rangle$, and therefore both the nonorthogonal overlap matrix and the orthogonalizing transformation are multiplicatively separable.
Specifically, $\mathbf{S}(\mu,\nu)=\mathbf{S}_{1}(\mu,\nu)\otimes\mathbf{S}_{2}(\mu,\nu)$ and $\mathbf{X}(\mu,\nu)=\mathbf{X}_{1}(\mu,\nu)\otimes\mathbf{X}_{2}(\mu,\nu)$.
Separability of the dressed states and parent Hamiltonian further ensures that the nonorthogonal effective Hamiltonian is additively separable, $\mathbf{H}(\mu,\nu)=\mathbf{H}_{1}(\mu,\nu)\otimes\mathbf{S}_{2}(\mu,\nu)+\mathbf{S}_{1}(\mu,\nu)\otimes\mathbf{H}_{2}(\mu,\nu)$.
Inserting these relationships into Eq.~\ref{eq:Ht} produces $\tilde{\mathbf{H}}(\mu,\nu)=\tilde{\mathbf{H}}_{1}(\mu,\nu)\otimes\mathbf{I}_{2}+\mathbf{I}_{1}\otimes\tilde{\mathbf{H}}_{2}(\mu,\nu)$, implying order-by-order size consistency as desired.

We now determine which linear combination of the perturbative coefficients of Eq.~\ref{eq:Ht} defines the truncation at a given order.
We require this truncation to preserve Hermiticity, so any retained bra- and ket-conjugate coefficients must enter with equal weight.
We also require that, in the single-reference limit, the effective Hamiltonian recover the parent RSPT energy at the same order.
In this limit, Eq.~\ref{eq:Ht} reduces to the Rayleigh quotient
\begin{align}
    \tilde{H}(\mu,\nu)
    =
    \frac{
    \langle A(\mu)\vert\hat{H}\vert A(\nu)\rangle
    }{
    \langle A(\mu)\vert A(\nu)\rangle
    }.
    \label{eq:Ht_sr}
\end{align}
Assuming that the single-reference RSPT used to obtain the wavefunction corrections invoked a reference-dependent partitioning, $\hat{H}(\lambda)=\hat{F}_{A}+\lambda\hat{W}_{A}$, the perturbative approximation of Eq.~\ref{eq:Ht_sr} through order $m$ is
\begin{align}
    \tilde{H}^{[m]}= \sum_{i+j<m}\tilde{H}^{(i,j)} +\sum_{i+j=m}\tilde{F}_{A}^{(i,j)},\label{eq:HtN_sr}
\end{align}
where we have inserted the perturbative expansion of the Hamiltonian $\hat{H}(\lambda)$.
Here, $\tilde{F}_{A}^{(i,j)}$ are the perturbative coefficients of $\tilde{F}_{A}(\mu,\nu)=X^{\dagger}(\nu,\mu)F_{A}(\mu,\nu)X(\mu,\nu)$, where $F_A(\mu,\nu)=\langle A(\mu)\vert\hat F_A\vert A(\nu)\rangle$ involves matrix elements of the zeroth-order Hamiltonian.
In Eq.~\ref{eq:HtN_sr}, only these zeroth-order terms contribute for $i+j=m$ because the remaining first-order terms increase the total perturbative order by one.

A natural way to generalize Eq.~\ref{eq:HtN_sr} to a multireference form is to promote each coefficient to its matrix counterpart. 
For the Hamiltonian contributions, this amounts to the substitution $\tilde{H}^{(i,j)}\rightarrow\tilde{\mathbf{H}}^{(i,j)}$. 
However, for the reference-dependent terms, $\tilde{F}_{A}^{(i,j)}$, a similar substitution would require a consistent definition of the zeroth-order Hamiltonian partition across the multireference model space. 
To avoid this, we recognize that each coefficient in Eq.~\ref{eq:HtN_sr} contributes energy corrections from multiple orders, enabling different linear combinations of coefficients to all yield the same order-$m$ RSPT energy. 
Eq.~\ref{eq:HtN_sr} is then one member of a family of functionals that are defined by different linear combinations of perturbative coefficients.
In Appendix~\ref{app:energy_functionals}, we formulate a system of equations whose solutions are the weights that define these linear combinations.
We additionally show that, for all orders $m\geq 2$, solutions exist in which the explicitly reference-dependent terms, $\tilde{F}^{(i,j)}_{A}$, vanish, allowing a multireference generalization without defining a common zeroth-order Hamiltonian across the model space. 

For target energy orders $m=2n$ and $m=2n+1$, we further restrict the retained coefficients to those involving wavefunction corrections up to order $n$. In addition to being consistent with the $2n+1$ rule of perturbation theories, this also serves to avoid undue cost by eliminating terms whose energy contributions cancel in the single-reference limit.
Two members of this restricted family of single-reference functionals are
\begin{align}
    \tilde{H}^{[2n]} = \sum_{i,j=0}^{n-1}\tilde{H}^{(i,j)} +\frac{1}{2}\sum_{i=0}^{n-1}\bigg(\tilde{H}^{(n,i)} + \tilde{H}^{(i,n)}\bigg)\label{eq:Ht_noF_even}
\end{align}
at even orders, and
\begin{align}
    \tilde{H}^{[2n+1]} = \sum_{i,j=0}^{n}\tilde{H}^{(i,j)}\label{eq:Ht_noF}
\end{align}
at odd orders. 
Replacing each scalar coefficient $\tilde{H}^{(i,j)}$ with its matrix counterpart $\tilde{\mathbf H}^{(i,j)}$ yields the corresponding multireference generalizations at all orders.
Restricting ourselves to only first-order wavefunction corrections (i.e., $n=1$) gives rise to both second- and third-order theories.
At second order, the NOCI-RS2 Hamiltonian is
\begin{align}
    \tilde{\mathbf H}^{[2]} = \tilde{\mathbf H}^{(0,0)} +\frac{1}{2}\left(\tilde{\mathbf H}^{(1,0)} + \tilde{\mathbf H}^{(0,1)}\right)\label{eq:NOCI_RS2}
\end{align}
and at third order, the NOCI-RS3 Hamiltonian is
\begin{align}
    \tilde{\mathbf H}^{[3]}
    =
    \tilde{\mathbf H}^{(0,0)}
    +
    \tilde{\mathbf H}^{(1,0)}
    +
    \tilde{\mathbf H}^{(0,1)}
    +
    \tilde{\mathbf H}^{(1,1)}.\label{eq:NOCI_RS3}
\end{align}
As shown in Appendix~\ref{app:energy_functionals}, within the class of functionals considered here, the weights defining the single-reference functionals in Eqs.~\ref{eq:Ht_noF_even} and \ref{eq:Ht_noF} are unique only for $n=1$.
Indeed, while the multireference forms implied by Eqs.~\ref{eq:Ht_noF_even} and \ref{eq:Ht_noF} are reasonable, uniquely specifying the effective Hamiltonian at higher orders may require additional constraints. 
We leave an exploration of such higher-order theories to future work.

Additionally, the cancellations used to obtain Eqs.~\ref{eq:Ht_noF_even} and \ref{eq:Ht_noF} hold in the single-reference limit but do not generally extend to off-diagonal matrix elements between different reference determinants. 
The present approach therefore yields a particular multireference generalization that excludes off-diagonal couplings which might otherwise contribute at the nominal order.
For example, at third order, Eq.~\ref{eq:HtN_sr} contains $\tilde{H}^{(2,0)}$ and $\tilde{H}^{(0,2)}$, whose direct promotion would introduce second-order wavefunction contributions that are absent from Eq.~\ref{eq:NOCI_RS3}.
While including such couplings may be beneficial, excluding them here allows us to avoid defining a zeroth-order partition of the Hamiltonian in the multireference model space. 
Finally, in Appendix~\ref{app:working_equations}, we provide expressions for each contribution to Eqs.~\ref{eq:NOCI_RS2} and \ref{eq:NOCI_RS3} in terms of their respective unprojected quantities, the calculation of which has been extensively discussed in prior literature.\cite{yostMultireferencePerturbationMethod2013,yostSizeConsistentFormulations2016,yostEfficientImplementationNOCIMP22018,burtonGeneralizedNonorthogonalMatrix2021,burtonGeneralizedNonorthogonalMatrix2022}

\section{Results and Discussion}

We now proceed with a numerical assessment of the NOCI-RSPT framework through simulations involving its second- and third-order manifestations, NOCI-RS2 and NOCI-RS3.
When appropriate, we compare against select prior NOCI-MP2 methods, including the original $\Delta$SCF(2) method of Ref.~\citenum{yostMultireferencePerturbationMethod2013} (referred to as NOCI-MP2 v0) and the more recent NOCI-MP2 v1 and v2 methods of Ref.~\citenum{yostSizeConsistentFormulations2016}.
For both NOCI-RSPT and NOCI-MP2, we construct first-order corrections from the corresponding M\o ller--Plesset amplitudes, excluding any singles contributions.
We first demonstrate the robust size consistency realized by the NOCI-RSPT framework by calculating the size-consistency error for assemblies of multireference H$_2$ fragments, after which we simulate the dissociation of H$_2$ to demonstrate the role of model-space projection.
Finally, we explore the consequences of large perturbative amplitudes and their remediation through enlargement of the model space and the use of regularized perturbation theory.

Here, we obtain the relevant orbital coefficients with Q-Chem version 6.2.1\cite{epifanovskySoftwareFrontiersQuantum2021} and obtain MP2 dressings and reference FCI energies with PySCF version 2.14.0.\cite{sunPythonSimulationsChemistry2026}
The NOCI-MP2 and NOCI-RSPT methods are implemented in a custom in-house code, which also computes the $\sigma$-regularized MP amplitudes used below.
For all regularized calculations, we use $\sigma=1.0\,E_{\mathrm{h}}^{-1}$.
The size-consistency calculations use the 6-311G basis set,\cite{krishnanSelfconsistentMolecularOrbital1980} while the H$_2$ dissociation calculations use the cc-pVDZ basis set.\cite{dunningGaussianBasisSets1989} Unless we explicitly specify otherwise, NOCI-RS2 and NOCI-RS3 use model-space projection, which we take as a defining feature of these methods.

\begin{figure}
\centering
\includegraphics[scale=1]{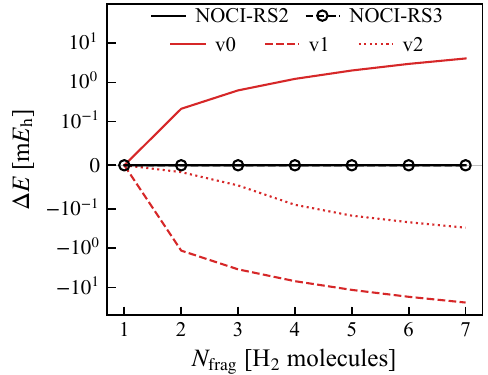}
\caption{Size-consistency error, $\Delta E=E_{N_{\mathrm{frag}}}-N_{\mathrm{frag}}E_1$, for NOCI-RSPT (NOCI-RS2 and NOCI-RS3) and NOCI-MP2 (v0, v1, and v2) for $N_{\mathrm{frag}}$ stretched H$_2$ molecules separated by $1000\,\AA$. Each H$_2$ molecule has a bond length of $1.5\,\AA$ and a local model space made up of two UHF determinants. All perturbative corrections are constructed from the associated MP amplitudes. Calculations were performed with the 6-311G basis set, and the vertical axis uses a linear region bounded by $\pm0.1\,\mathrm{m}E_{\mathrm{h}}$.}
\label{fig:FIG_SC}
\end{figure}

In Fig.~\ref{fig:FIG_SC}, we show the size-consistency errors obtained for each method applied to a system of up to seven H$_2$ molecules.
Here, each molecule has a bond length of $1.5\,\AA$ (i.e., stretched just beyond twice its equilibrium bond length), and adjacent molecules are separated by $1000\,\AA$ along a direction perpendicular to their internuclear axes.
This stretched bond length causes the mean-field determinants describing the ground state to split into two symmetry-broken unrestricted Hartree--Fock (UHF) determinants.
We use these two local spin configurations to enumerate the $N_{\mathrm{det}}=2^{N_{\mathrm{frag}}}$ UHF spin isomers that make up the model space of each assembly of $N_{\mathrm{frag}}$ H$_2$ molecules.
Each of these determinants corresponds to one of the ways of distributing the spin-up and spin-down electrons within each H$_2$ fragment.
Together, the determinants satisfy the model-space closure requirement necessary for size consistency in NOCI-RSPT.
As a result, NOCI-RS2 and NOCI-RS3 exhibit size-consistency errors on the order of numerical noise, whereas all NOCI-MP2 methods exhibit size-consistency errors that are considerably larger.

The size-consistency errors exhibited by NOCI-MP2 v1 and v2 may initially be surprising, given that the construction of these versions was motivated by a restoration of size consistency that was absent in v0.
This restoration, however, was limited to the case of a single multireference fragment and a single-reference environment.\cite{yostSizeConsistentFormulations2016}
NOCI-RSPT, on the other hand, realizes a more robust size consistency that extends to arbitrary multireference fragments.
We realize this broader size consistency as a direct result of the orthogonalization of the effective Hamiltonian and overlap matrices prior to perturbative truncation.
As explained in the previous section and demonstrated numerically in Fig.~\ref{fig:FIG_SC}, this ensures that any combination of the perturbative coefficients in Eq.~\ref{eq:Ht} yields a size-consistent theory.
While none of the NOCI-MP2 methods is generally size consistent, the limited notion of size consistency established for the v1 and v2 methods still enables them to be applied to a wide range of relevant multireference problems.
In subsequent numerical comparisons, we therefore exclude NOCI-MP2 v0, which is not size consistent even in these more limited cases.

\begin{figure}
    \centering
    \includegraphics[scale=1]{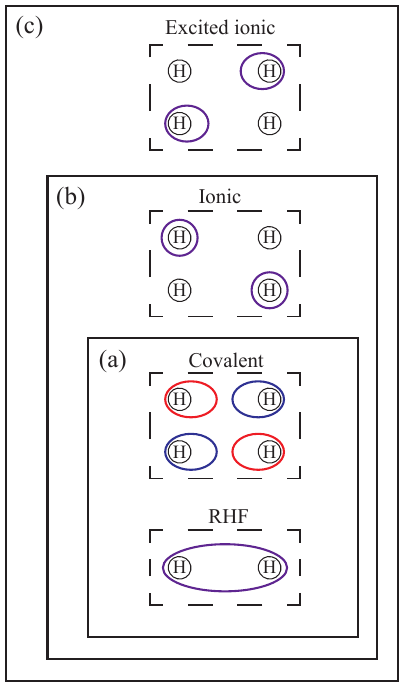}
    \caption{Schematic depiction of occupied orbitals used to construct three progressively larger model spaces. Doubly occupied orbitals are depicted in purple, while singly occupied orbitals are depicted in red (spin up) or blue (spin down). Boxes (a), (b), and (c) contain the three-determinant, five-determinant, and seven-determinant NOCI model spaces considered here.}
    \label{fig:FIG_DET}
\end{figure}

In addition to the challenge of realizing a size-consistent theory, another persistent challenge in the application of methods based on nonorthogonal determinants is the selection of an appropriate model space.
Indeed, one may initially consider the two symmetry-broken UHF determinants (i.e., those used for the single-molecule case in Fig.~\ref{fig:FIG_SC}) as a model space to describe H$_2$ across its dissociation curve.
While these may provide a good approximation of the system in the dissociative limit, when the internuclear distance is reduced past the Coulson--Fischer (CF) point, the two UHF determinants merge into a single restricted Hartree--Fock (RHF) determinant.\cite{coulsonXXXIVNotesMolecular1949}
The sudden reduction in the model-space size will lead to a kink in the potential energy surface and a reduction to the single-reference energy near equilibrium.
Instead, we can construct a similar model space in which all the determinants remain well defined across the dissociation curve by employing a constrained relaxation approach known as the method of absolutely localized molecular orbitals (ALMO).\cite{khaliullinEfficientSelfconsistentField2006,hornUnrestrictedAbsolutelyLocalized2013}
Within the ALMO approach, each hydrogen atom is treated as a fragment on which each electron remains fully localized.
Thus, one orbital is optimized from atomic orbitals of one hydrogen, and the other orbital of opposite spin is optimized from atomic orbitals of the other hydrogen. 
Consequently, the symmetry-broken UHF-like determinants remain well defined even when the internuclear distance crosses the CF point.
Near equilibrium, however, these covalent determinants increase in energy, and the RHF determinant becomes a better descriptor.
We therefore combine the two covalent determinants with the RHF determinant, resulting in a three-determinant model space that is depicted schematically in Fig.~\ref{fig:FIG_DET}(a).
This model space also resembles those that may arise in molecular biradicals, where two localized open-shell configurations describe the dominant static correlation when the radical centers are well separated, while closed-shell configurations become increasingly important as they interact more strongly.
The H$_2$ dissociation problem therefore provides a simple setting in which to examine features of NOCI-RSPT methods that may also be relevant to this broader class of chemical systems.

\begin{figure}
    \centering
    \includegraphics[scale=1]{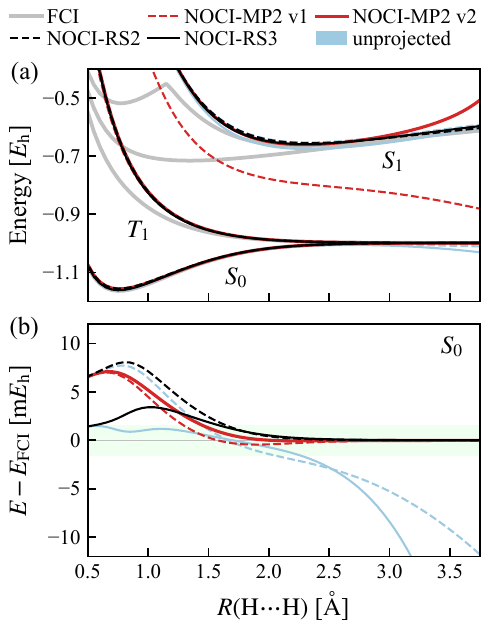}
    \caption{Low-energy spectrum of H$_2$ as a function of bond length (a) and corresponding ground-state energy errors (b) using the three-determinant model space depicted in Fig.~\ref{fig:FIG_DET}(a). Light blue traces correspond to NOCI-RSPT results obtained without model-space projection, and the green shaded region corresponds to chemical accuracy at $\pm 1.0\,\mathrm{kcal}/\mathrm{mol}$. Calculations utilized the cc-pVDZ basis set.}
    \label{fig:FIG_PRO}
\end{figure}

In Fig.~\ref{fig:FIG_PRO}(a), we apply the three-determinant model space to the calculation of the low-energy spectrum of H$_2$ using NOCI-RS2, NOCI-RS3, and NOCI-MP2 v1 and v2.
Because the model space consists of only three determinants, the eigenstates of the effective Hamiltonian can only describe three of the energy levels.
This particular choice of model space enables us to capture the ground-state singlet as well as a triplet excited state and a singlet excited state.
The triplet state becomes degenerate with the singlet ground state at dissociation, while the higher-lying singlet state remains well separated in energy.
Each method exhibits qualitative agreement with FCI for the first (triplet) excited state.
For the singlet excited state, however, both NOCI-RSPT methods exhibit qualitative agreement, while their unprojected counterparts exhibit a small deviation.
NOCI-MP2 v1 and v2, however, exhibit significant qualitative inaccuracies for this excited state.
NOCI-MP2 v1, in particular, exhibits deviations on the order of hundreds of millihartrees across the dissociation curve.
NOCI-MP2 v2 exhibits more modest errors, primarily deviating from the FCI result at dissociation.

In Fig.~\ref{fig:FIG_PRO}(b), we highlight the deviation of the ground-state potential energy surface from FCI.
In the dissociative limit, all methods except the unprojected NOCI-RSPT methods converge to the FCI energy.
While all of the NOCI-MP2 and NOCI-RSPT methods are nonvariational, only the unprojected NOCI-RSPT methods and NOCI-MP2 v1 exhibit energies below the FCI value in this case.
Interestingly, unprojected NOCI-RS3 exhibits a larger deviation than its second-order counterpart at dissociation, in contrast to the near-equilibrium regime, where the third-order contribution improves the accuracy over both NOCI-RS2 and the NOCI-MP2 methods.

It is important to recognize at this point that neither NOCI-MP2 v1 nor v2 incorporates model-space projection in its construction.
Despite this, they generally exhibit better ground-state accuracy across the dissociation curve than the unprojected NOCI-RS2 method, demonstrating that the consistent perturbative formulation realized in NOCI-RSPT in isolation does not always lead to improved accuracy.
Indeed, it is the combination of model-space projection and consistent perturbative truncation that endows NOCI-RSPT with more robust performance.
This can be understood by recognizing that inserting the projected dressed states (Eq.~\ref{eq:ket}) into the construction of NOCI-MP2 v0, v1, and v2 causes all three methods to reduce to NOCI-RS2.
We demonstrate this reduction algebraically in Appendix~\ref{app:reduction}.
NOCI-RSPT therefore provides a systematic framework in which model-space projection and consistent perturbative ordering yield a common second-order theory and permit its extension to higher orders.

The performance of multireference theories can be severely degraded by the presence of so-called intruder states.\cite{choeIdentifyingRemovingIntruder2001}
In a perturb-then-diagonalize approach, these states arise when the single-reference theory applied to a model-space determinant produces large wavefunction corrections due to vanishing orbital-energy gaps appearing in the denominators of perturbative amplitudes.
This can lead to large perturbative amplitudes and, in particularly severe cases, divergences that dramatically disrupt the accuracy of a multireference method.
One approach for mitigating intruder-state effects is to expand the model space to include the problematic configurations.\cite{burtonReachingFullCorrelation2020,pulayPerspectiveCASPT2Method2011}
This is motivated by the notion that such determinants can be better incorporated via mixing within the model space than through the perturbative dressing.
One can reasonably expect that such an approach would only work if the multireference theory incorporates model-space projection.
If not, the inclusion of the intruder state in the model space may result in even worse performance because the excitation remains in the wavefunction correction while also being redundant with a model-space determinant.

Intruder-state pathology can manifest in NOCI-RSPT and NOCI-MP2 methods when the model space is expanded in an effort to improve their performance.
Indeed, motivated by the prospect of improving the near-equilibrium accuracy, one may expand the model space to include two ``ionic'' determinants that each localize both electrons on an individual hydrogen atom.
These configurations become important when the two hydrogen atoms are close together but can lead to intruder states as the internuclear distance grows.
We depict two such ionic configurations in Fig.~\ref{fig:FIG_DET}(b), where relaxed orbitals are obtained by using the ALMO approach to relax doubly occupied fragment-local orbitals (i.e., $\mathrm{H}^{+}\,\mathrm{H}^{-}$ and $\mathrm{H}^{-}\,\mathrm{H}^{+}$).

\begin{figure}
    \centering
    \includegraphics[scale=1]{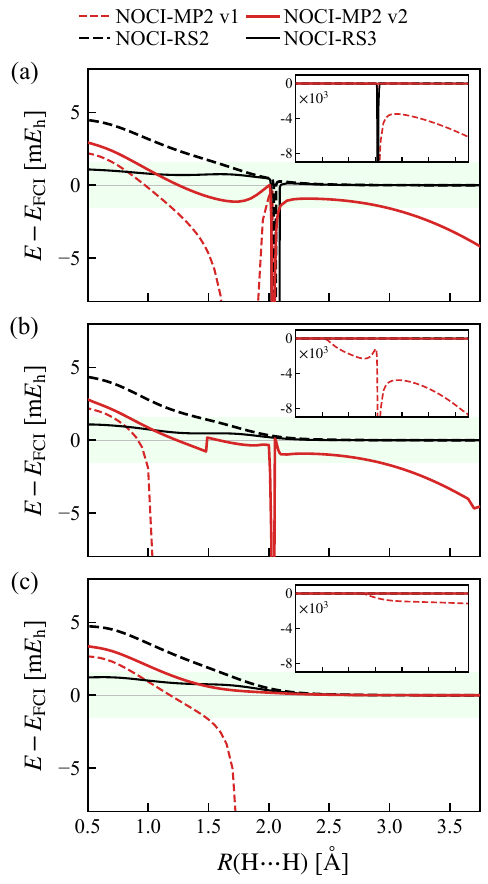}
    \caption{Ground-state energy errors obtained along the dissociation curve of H$_2$ for (a) the five-determinant model space of Fig.~\ref{fig:FIG_DET}(b), (b) the seven-determinant model space of Fig.~\ref{fig:FIG_DET}(c), and (c) the five-determinant model space using $\sigma$-regularized MP amplitudes. Insets show a zoomed-out perspective of the same curves. Green shaded regions correspond to chemical accuracy of $\pm 1.0\,\mathrm{kcal}/\mathrm{mol}$. Calculations utilized the cc-pVDZ basis set.}
    \label{fig:FIG_MSR}
\end{figure}

The ground-state energy errors for the five-determinant model space are shown in Fig.~\ref{fig:FIG_MSR}(a), where all NOCI-RSPT and NOCI-MP2 methods studied exhibit a divergence near $2.0\,\AA$.
At shorter separations, electrostatic interactions stabilize the doubly occupied 1s-like orbital on the $\mathrm{H}^{-}$ fragment, lowering its energy below that of the unoccupied 1s-like orbital on the $\mathrm{H}^{+}$ fragment and producing a finite orbital-energy gap.
As the internuclear distance increases, these interactions weaken while the electron repulsion remains, causing the gap to close near $2.0\,\AA$ and producing a pole in the MP amplitudes.
Beyond this crossing, the occupied orbital lies above the unoccupied orbital, rendering the ionic determinants non-Aufbau.
Notably, the resulting energy error is on the scale of tens of millihartrees in NOCI-MP2 v2 and NOCI-RS2 but extends to the scale of hartrees in NOCI-MP2 v1 and NOCI-RS3.
The more severe inaccuracy in NOCI-RS3 is consistent with a breakdown of the perturbative description due to the large magnitude of the wavefunction correction.
In this regime, the additional contributions retained in the third-order theory amplify the errors associated with the large amplitudes.
Indeed, model-space projection cannot fully remediate this behavior because the problematic terms correspond to an electronic configuration that lies only partially within the model space.
Instead, model-space projection serves to improve the behavior in the vicinity of the divergence by reducing the sensitivity of the energy to the divergence while not completely removing it.
Ultimately, this leads to a dramatic improvement in the accuracy of NOCI-RS3, which achieves errors below the chemical accuracy threshold except in the close vicinity of the divergence.

To remove the divergence, one can consider including the problematic electronic configuration in the model space, after which model-space projection will remove its divergent amplitude from the wavefunction correction.
Indeed, because the divergence in this case is associated with a single amplitude arising from each ionic determinant's first-order correction, the model space only needs to be expanded to include two additional configurations.
These configurations are depicted schematically in Fig.~\ref{fig:FIG_DET}(c) and correspond to double excitations of each ionic determinant that transfer both electrons onto the opposing hydrogen atom.
We leave the orbitals comprising these configurations unrelaxed to ensure that the model-space projection fully eliminates the diverging amplitudes.

The result of using this expanded seven-determinant model space is shown in Fig.~\ref{fig:FIG_MSR}(b), where NOCI-RS2 and NOCI-RS3 are now entirely divergence-free, with NOCI-RS3 exhibiting accuracy to within $1.2\, \mathrm{m}E_{\mathrm{h}}$ across the entire dissociation curve.
In contrast, the divergence remains in both NOCI-MP2 v1 and v2 because neither incorporates model-space projection.
NOCI-MP2 v1, in particular, exhibits an even more rapid deviation below the FCI energy for internuclear distances below $2.0\,\AA$ and worse overall accuracy, with errors on the order of hartrees.
As in the case of the five-determinant model space (Fig.~\ref{fig:FIG_MSR}(a)), NOCI-MP2 v2 exhibits a smaller divergence than NOCI-MP2 v1 at $2.0\,\AA$.
However, with the additional excited ionic determinants, a new discontinuity is introduced near $1.5\,\AA$.
This is not associated with an intruder state but is instead due to numerical instability associated with the overlap-matrix construction in NOCI-MP2 v2.
While the disparity between NOCI-MP2 and NOCI-RSPT methods is indeed stark, it is ultimately unsurprising since neither NOCI-MP2 method incorporates model-space projection.

An alternative approach that does not rely on model-space projection is to regularize the perturbative amplitudes directly.
Indeed, prior work on NOCI-MP2 used substantial level shifts to tame poorly behaved amplitudes.\cite{yostMultireferencePerturbationMethod2013}
Here, we consider $\sigma$-regularized MP amplitudes, in which the regularizer depends on the energy denominator entering the first-order amplitudes.\cite{sheeRegularizedSecondOrderMoller2021,leeRegularizedOrbitalOptimizedSecondOrder2018}
Importantly, using regularized amplitudes does not guarantee that NOCI-RSPT (or NOCI-MP2) reduces to the associated regularized perturbation theory in the single-reference limit, since regularization approaches typically modify the energy functional.\cite{leeRegularizedOrbitalOptimizedSecondOrder2018}
For NOCI-RS2 with $\sigma$-regularized amplitudes, the single-reference limit is the $\sigma$-MP2 energy, while for NOCI-RS3, using the same amplitudes amounts to evaluating the third-order functional in Eq.~\ref{eq:NOCI_RS3} in its single-reference form.
To our knowledge, this does not correspond to a previously formulated third-order regularized single-reference perturbation theory.
We use $\sigma=1.0\,E_{\mathrm{h}}^{-1}$ throughout, whereas Ref.~\citenum{sheeRegularizedSecondOrderMoller2021} recommended $\sigma=0.7\,E_{\mathrm{h}}^{-1}$ for general use.
The latter produces stronger regularization than required in the present case.

In Fig.~\ref{fig:FIG_MSR}(c), we show the results obtained using the $\sigma$-regularized MP amplitudes.
Regularization removes the divergence in both NOCI-RSPT and NOCI-MP2 methods.
Near equilibrium, the regularized amplitudes cause a slightly larger energy error due to the reduced portion of the correlation energy captured by the regularized wavefunction corrections.
This source of error could be reduced by choosing a larger value of $\sigma$. As in the seven-determinant case, NOCI-RS3 exhibits errors within chemical accuracy (less than $1.0\,\mathrm{kcal}/\mathrm{mol}$ error) across the dissociation curve.
NOCI-MP2 v2 exhibits improved smoothness across its potential energy surface, particularly in the dissociation limit, where it no longer goes below the FCI energy.
Likewise, NOCI-MP2 v1 exhibits smaller inaccuracies and smoother behavior but still deviates from FCI on the hartree scale.
Clearly, the regularity of amplitudes alone is not sufficient to ensure the accuracy of NOCI-MP2 methods.

We conclude this section with a brief discussion of the cost scaling of the NOCI-RS2 and NOCI-RS3 methods.
For a fixed model-space size, NOCI-RS2, like NOCI-MP2, scales as $O(N_{\mathrm{orb}}^{5})$ in the number of orbitals $N_{\mathrm{orb}}$, matching the cost scaling of MP2.
This is straightforward to see from the working expressions detailed in Appendix~\ref{app:working_equations}, which demonstrate that the NOCI-RS2 Hamiltonian can be constructed from exactly the same terms required to construct the NOCI-MP2 methods.
The additional terms required for NOCI-RS3 involve Hamiltonian and overlap matrix elements between first-order wavefunction corrections on both the bra and ket sides.
These terms can be evaluated at $O(N_{\mathrm{orb}}^{6})$ cost, matching the cost scaling of MP3.
Model-space projection requires no additional orbital-index contractions in the NOCI-RSPT methods.
The model-space projection, orthogonalization, and diagonalization scale as $O(N_{\mathrm{det}}^{3})$ with the number $N_{\mathrm{det}}$ of model-space determinants.
Including the evaluation of all determinant pairs, the overall scaling of NOCI-RS2 is therefore $O(N_{\mathrm{det}}^{3}+N_{\mathrm{det}}^{2}N_{\mathrm{orb}}^{5})$, and the overall scaling of NOCI-RS3 is $O(N_{\mathrm{det}}^{3}+N_{\mathrm{det}}^{2}N_{\mathrm{orb}}^{6})$.
While we have not discussed the memory scaling of the methods here, the RI approximation has been used to improve the efficiency of NOCI-MP2 methods, and similar techniques would certainly be applicable here.\cite{yostEfficientImplementationNOCIMP22018}
Similarly, recent work has established the potential of local correlation approaches to yield reduced-scaling and even linear-scaling implementations of single-reference perturbation theories.\cite{maslenNoniterativeLocalSecond1998,schutzLoworderScalingLocal1999,ayalaLinearScalingSecondorder1999,shiLocalSecondOrder2024,wangMoreNumericalPrecision2025,wangRegularizedSecondOrderMoller2025a,wangLocalSecondOrderMoller2023a}
While the same sparsity assumptions\cite{pinskiSparseMapsSystematic2015,wangSparsityElectronRepulsion2023} would have to be extended to the off-diagonal terms, it is possible that such approaches could enable a dramatic reduction in the cost scaling of NOCI-RSPT methods as well.

\section{Conclusion}

We have introduced NOCI-RSPT, a framework for constructing perturb-then-diagonalize multireference perturbation theories from well-established single-reference methods.
NOCI-RSPT realizes robust size consistency, eliminates model-space redundancies, and admits a systematic extension to higher perturbative orders.
It does so by orthogonalizing the effective Hamiltonian prior to perturbative truncation, incorporating model-space projection at the level of the wavefunction ansatz, and using constraints derived from the single-reference limit to guide the construction of the multireference effective Hamiltonian.

Our numerical results confirm robust size consistency and demonstrate the importance of combining consistent perturbative ordering with model-space projection.
For H$_2$, this combination yields more reliable ground- and excited-state energies than prior NOCI-MP2 methods, which incorporate neither feature.
Model-space expansion and amplitude regularization provide complementary strategies for controlling large perturbative corrections, including those associated with intruder states.
With either the expanded model space or the regularized amplitudes, NOCI-RS3 achieves errors within chemical accuracy across the H$_2$ dissociation curve, further demonstrating the utility of extending the NOCI-RSPT construction beyond second order.

The compact model spaces considered for H$_2$ also closely resemble those that can arise in molecular biradicals, where a small number of localized open-shell and closed-shell determinants capture the dominant static-correlation structure.
The results here therefore suggest that NOCI-RSPT may provide a promising low-cost treatment of both static and dynamic correlation in biradicals and related systems that admit compact nonorthogonal model spaces.
Establishing how broadly these conclusions extend beyond H$_2$ will require calculations on chemically realistic biradicals and other multireference systems.

For a fixed model-space size, NOCI-RS2 and NOCI-RS3 retain the orbital scaling of MP2 and MP3, respectively, although systematic construction of compact nonorthogonal model spaces remains an outstanding challenge.
The improved numerical stability demonstrated here may broaden the range of such model spaces that can be explored.
Broader molecular benchmarks and systematic strategies for constructing model spaces using related approaches, including resonating Hartree--Fock\cite{fukutomeTheoryResonatingQuantum1988} and more recent tensor-optimization methods,\cite{giulianiPreciseQuantumChemistry2026,giulianiVariationalLowenergySubspaces2026} are natural next steps toward establishing the practical scope of NOCI-RSPT.

\begin{acknowledgments}
A.K. thanks Wendy Billings and Dr. Rebecca Tomann for helpful discussions on the preparation of the ALMO states and Dr. Avijit Shee for discussions on the properties of multireference perturbation theories.
The authors acknowledge the use of Anthropic's Claude (Opus and Fable models) for assistance with the computational implementation and OpenAI's ChatGPT (Sol and Astra models) for assistance with analytical derivations and manuscript preparation.
This work was supported by the U.S. Department of Energy, Office of Science, Office of Basic Energy Sciences, through the Ultrafast Molecular Optical Sciences Program, under Contract DE-AC02-05CH11231. 
This research used resources of the National Energy Research Scientific Computing Center (NERSC), a U.S. Department of Energy Office of Science User Facility located at Lawrence Berkeley National Laboratory, operated under Contract DE-AC02-05CH11231.
\end{acknowledgments}

\section*{Author Declarations}

M.H.G. is a part-owner of Q-Chem Inc, which is the software
platform used to obtain inputs for the algorithms developed and assessed here.

\section*{Data Availability}

The data that support the findings of this study are available from the corresponding author upon reasonable request.

\appendix

\section{Construction of a reference-independent functional}
\label{app:energy_functionals}

In this Appendix, we derive a system of linear equations whose solutions are weights that define an order-$m$ RSPT energy functional.
Introducing real-valued weights into Eq.~\ref{eq:HtN_sr} and equating the resulting expression to the order-$m$ RSPT energy gives
\begin{align}
    \sum_{i+j< m}h_{ij}\tilde{H}^{(i,j)} +\sum_{i+j=m}f_{ij}\tilde{F}_{A}^{(i,j)} =\sum_{k=0}^{m}E^{(k)},\label{eq:e_crit}
\end{align}
where $E^{(k)}$ is the order-$k$ energy correction.

For the partitioning $\hat{H}(\lambda)=\hat{F}_{A}+\lambda\hat{W}_{A}$, the wavefunction corrections satisfy the RSPT amplitude equations
\begin{align}
    \hat{F}_{A}\vert A^{(r)}\rangle +\hat{W}_{A}\vert A^{(r-1)}\rangle = \sum_{k=0}^{r}E^{(k)}\vert A^{(r-k)}\rangle.\label{eq:rspt_amplitudes}
\end{align}
After left-multiplying by $\langle A(\mu)\vert \nu^{r}/\langle A(\mu)\vert A(\nu)\rangle$ and summing over $r$, we obtain
\begin{align}
    (1-\nu)\tilde{F}_{A}(\mu,\nu)+\nu\tilde{H}(\mu,\nu)=E(\nu).\label{eq:pert_1}
\end{align}
We also obtain the conjugate expression $(1-\mu)\tilde{F}_{A}(\mu,\nu)+\mu\tilde{H}(\mu,\nu)=E(\mu)$ by exchanging $\mu$ and $\nu$ in Eq.~\ref{eq:pert_1} and conjugating the result.
Subtracting these expressions yields the equality
\begin{align}
    \tilde{H}(\mu,\nu)-\tilde{F}_{A}(\mu,\nu)=\frac{E(\nu)-E(\mu)}{\nu-\mu},
\end{align}
where equating coefficients of $\mu^{i}\nu^{j}$ on both sides gives the equalities $\tilde{H}^{(i,j)}-\tilde{F}_{A}^{(i,j)}=E^{(i+j+1)}$ and therefore  $\tilde{H}(\mu,\nu)-\tilde{F}_{A}(\mu,\nu)=\sum_{i,j=0}\mu^{i}\nu^{j}E^{(i+j+1)}$.
Inserting this expression into Eq.~\ref{eq:pert_1} and collecting coefficients of $\mu^{i}\nu^{j}$ yields
\begin{align}
    \tilde F_A^{(i,j)}
    =
    \delta_{i0}\delta_{j0}E^{(0)}
    -(1-\delta_{i0})(1-\delta_{j0})E^{(i+j)},\label{eq:F_sr}
\end{align}
and subsequently
\begin{align}
    \tilde H^{(i,j)}
    &=\tilde F_A^{(i,j)}+E^{(i+j+1)}.
    \label{eq:H_sr}
\end{align}

Inserting Eqs.~\ref{eq:F_sr} and \ref{eq:H_sr} into Eq.~\ref{eq:e_crit} and collecting terms for each value of $k$ defines a system of equations in terms of the weights $h_{ij}$ and $f_{ij}$. 
The solutions of these equations, when inserted into the left-hand side of Eq.~\ref{eq:e_crit}, define an order-$m$ RSPT energy functional.
As described in the main text, we enforce three requirements on the weights that constitute an acceptable single-reference functional.  
First, Hermiticity of the corresponding multireference form requires that $h_{ij}=h_{ji}$ and $f_{ij}=f_{ji}$.
Second, reference independence requires that $f_{ij}=0$.
Third, including only wavefunction corrections up to order $n$ (for energy orders $m=2n$ and $m=2n+1$) requires that all weights where $i>n$ or $j>n$ vanish. 
With these requirements, the only terms entering into Eq.~\ref{eq:e_crit} are  
\begin{widetext}
\begin{align}
    h_{ij}\tilde{H}^{(i,j)} :
    \begin{pmatrix}
        h_{00}(E^{(0)}+E^{(1)})
        & h_{01}E^{(2)}
        & \cdots
        & h_{0n}E^{(n+1)} \\
        h_{10}E^{(2)}
        & h_{11}(E^{(3)}-E^{(2)})
        &
        & h_{1n}(E^{(n+2)}-E^{(n+1)}) \\
        \vdots & & \ddots & \vdots \\
        h_{n0}E^{(n+1)}
        & h_{n1}(E^{(n+2)}-E^{(n+1)})
        & \cdots
        & h_{nn}(E^{(2n+1)}-E^{(2n)})
    \end{pmatrix}.
    \label{eq:Hij}
\end{align}
\end{widetext}

Inspecting Eq.~\ref{eq:Hij} reveals that $h_{00}=1$ is required to recover the zeroth- and first-order energy terms, and that for values of $2\leq k\leq m-1$
\begin{align}
    \sum_{i+j=k-1}h_{ij} - \sum_{\substack{i+j=k\\ i,j\geq 1}}h_{ij}=1,\label{eq:req_1}
\end{align}
where all weights with $i>n$ or $j>n$ are zero.

For odd order $m=2n+1$, the constraint on the final term $k=m$ is $h_{nn}=1$, and Eq.~\ref{eq:req_1} can be satisfied by setting all nonzero weights to unity, which corresponds to the functional in Eq.~\ref{eq:Ht_noF}.
At even order $m=2n$, $h_{nn}=0$ and the $k=m$ case constrains $h_{n-1,n}+h_{n,n-1}=1$. 
Setting $h_{nn}=0$ in Eq.~\ref{eq:Hij} and taking all other weights to be unity reveals an extra contribution of $E^{(2n)}$.
Summing the last row and column (excluding the corner) yields $2 E^{(2n)}$, suggesting that the surplus contribution can be eliminated by taking $h_{in}=h_{ni}=\frac{1}{2}$ for $i<n$, also satisfying the $k=m$ constraint.  
The even-order solution is then $h_{in}=h_{ni}=\frac{1}{2}$ for $i<n$, $h_{ij}=1$ for $i,j<n$, and $h_{nn}=0$, which corresponds to the functional in Eq.~\ref{eq:Ht_noF_even}.

To demonstrate that these solutions are unique for $n=1$, we consider the terms that comprise the top-left entries of Eq.~\ref{eq:Hij} for $i,j\leq 1$.
For $m=2$, Eq.~\ref{eq:req_1} vanishes and the remaining constraint is $h_{01} + h_{10} = 1$. Hermiticity makes the solution $h_{01}=\frac{1}{2}$ unique.
For $m=3$, Eq.~\ref{eq:req_1} remains only for $k=2$, yielding
\begin{align}
    h_{01}+h_{10}-h_{11} = 1.
\end{align}
When combined with the constraint $h_{11}=1$, the only solution satisfying Hermiticity is $h_{01}=h_{10}=1$.

Finally, we show that the stated requirements do not uniquely determine the weights for energy orders $m\geq 4$ (i.e., wavefunction orders $n\geq 2$).
Eqs.~\ref{eq:F_sr} and \ref{eq:H_sr} imply the identity
\begin{align}
    \tilde H^{(2,0)}+\tilde H^{(0,2)}
    -2\tilde H^{(1,1)}
    -\tilde H^{(1,0)}-\tilde H^{(0,1)}=0.
\end{align}
Adding a real multiple of this to any admissible functional alters its corresponding weights without changing the RSPT energy or violating any of the requirements of a solution.
Thus, for $n\geq 2$, the solution for the weights is non-unique. 

\section{Working Expressions}
\label{app:working_equations}

For NOCI-RS2 and NOCI-RS3, only the first-order wavefunction corrections are required.
Expanding Eq.~\ref{eq:ket} to first order yields the projected correction
\begin{align}
    \vert A^{(1)}\rangle &= (1-\hat{P})\hat{\Omega}_{A}^{(1)}\vert A^{(0)}\rangle,
\end{align}
whose unprojected counterpart can be obtained by setting $\hat{P}\rightarrow 0$.
The coefficients of the orthogonalizing transformation are
\begin{align}
    \mathbf{X}^{(0,0)} &= \big(\mathbf{S}^{(0,0)}\big)^{-\frac{1}{2}},\\
    \mathbf{X}^{(1,0)} &= -\frac{1}{2}\mathbf{X}^{(0,0)}\mathbf{G}^{(1,0)},\\
    \mathbf{X}^{(0,1)} &= -\frac{1}{2}\mathbf{X}^{(0,0)}\mathbf{G}^{(0,1)},\\
    \mathbf{X}^{(1,1)} &= \mathbf{X}^{(0,0)}\bigg[-\frac{1}{2} \mathbf{G}^{(1,1)}+\frac{3}{8}\big\{\mathbf{G}^{(1,0)},\mathbf{G}^{(0,1)}\big\}\bigg],
\end{align}
where we have expanded $(\mathbf{G}+\mathbf{I})^{-\frac{1}{2}}$ using the binomial series and taken $\mathbf{G}^{(i,j)}=\big(\mathbf{S}^{(0,0)}\big)^{-\frac{1}{2}}\mathbf{S}^{(i,j)}\big(\mathbf{S}^{(0,0)}\big)^{-\frac{1}{2}}$ for $i+j>0$.
For compactness, we define the anticommutator $\{\mathbf{A},\mathbf{B}\}=\mathbf{A}\mathbf{B}+\mathbf{B}\mathbf{A}$.

The coefficients of the orthogonalized Hamiltonian are generally given by
\begin{align}
    \tilde{\mathbf{H}}^{(i,j)} = \sum_{\substack{a+c+e=i\\b+d+f=j}}\big(\mathbf{X}^{(b,a)}\big)^{\dagger}\mathbf{H}^{(c,d)}\mathbf{X}^{(e,f)},
\end{align}
where all summation indices are nonnegative integers.
The overlap and nonorthogonal Hamiltonian coefficients follow by expanding Eqs.~\ref{eq:S_AB} and \ref{eq:H_AB}, respectively:
\begin{align*}
    \big[\mathbf S^{(i,j)}\big]_{AB} &= \langle A^{(i)}\vert B^{(j)}\rangle,\\
    \big[\mathbf H^{(i,j)}\big]_{AB} &= \langle A^{(i)}\vert\hat H\vert B^{(j)}\rangle.
\end{align*}

Up to this point, the expressions in this appendix are applicable to both projected and unprojected dressed states.
Invoking the model-space intermediate normalization enforced by the projection reduces many terms to zero.
In particular, $\mathbf{X}^{(0,i)}=\mathbf{X}^{(i,0)}=0$ for $i>0$.
The nonzero contributions needed for NOCI-RS2 and NOCI-RS3 are
\begin{align}
    \tilde{\mathbf{H}}^{(0,0)}&=\big(\mathbf{X}^{(0,0)}\big)^{\dagger}\mathbf{H}^{(0,0)}\mathbf{X}^{(0,0)}, \\
    \tilde{\mathbf{H}}^{(1,0)}&=\big(\mathbf{X}^{(0,0)}\big)^{\dagger}\mathbf{H}^{(1,0)}\mathbf{X}^{(0,0)}, \\
    \tilde{\mathbf{H}}^{(0,1)}&=\big(\mathbf{X}^{(0,0)}\big)^{\dagger}\mathbf{H}^{(0,1)}\mathbf{X}^{(0,0)}, \\
    \tilde{\mathbf{H}}^{(1,1)}&=\big(\mathbf{X}^{(0,0)}\big)^{\dagger}\mathbf{H}^{(1,1)}\mathbf{X}^{(0,0)} \\
    &- \frac{1}{2}\bigg\{\big(\mathbf{X}^{(0,0)}\big)^{\dagger}\mathbf{S}^{(1,1)}\mathbf{X}^{(0,0)},\tilde{\mathbf{H}}^{(0,0)}\bigg\}. \nonumber
\end{align}

Inserting these definitions into Eqs.~\ref{eq:NOCI_RS2} and \ref{eq:NOCI_RS3} from the main text yields
\begin{align}
    \tilde{\mathbf{H}}^{[2]} = \big(\mathbf{X}^{(0,0)}\big)^{\dagger}\bigg[\mathbf{H}^{(0,0)}+\frac{1}{2}\bigg(
    \mathbf{H}^{(1,0)}+\mathbf{H}^{(0,1)}\bigg) \bigg]\mathbf{X}^{(0,0)}\label{eq:NOCI_RS2_full}
\end{align}
and
\begin{align}
    \tilde{\mathbf{H}}^{[3]} &= \big(\mathbf{X}^{(0,0)}\big)^{\dagger}\bigg[\mathbf{H}^{(0,0)}+
    \mathbf{H}^{(1,0)}+\mathbf{H}^{(0,1)} + \mathbf{H}^{(1,1)}\label{eq:NOCI_RS3_full}\\
    &-\frac{1}{2}\bigg(\mathbf{S}^{(1,1)}(\mathbf{S}^{(0,0)})^{-1}\mathbf{H}^{(0,0)}+\mathbf{H}^{(0,0)}(\mathbf{S}^{(0,0)})^{-1}\mathbf{S}^{(1,1)}\bigg) \bigg]\mathbf{X}^{(0,0)}.\nonumber
\end{align}

We can compute each of the terms in Eqs.~\ref{eq:NOCI_RS2_full} and \ref{eq:NOCI_RS3_full} using terms obtained with the unprojected wavefunction corrections.
In the following expressions, we denote unprojected quantities by the subscript $\mathrm u$.
The projected matrix elements of the Hamiltonian are
\begin{align}
\bigl[\mathbf H^{(0,0)}\bigr]_{AB}
&= \bigl[\mathbf H_{\mathrm u}^{(0,0)}\bigr]_{AB},
\end{align}
\begin{align}
\bigl[\mathbf H^{(1,0)}\bigr]_{AB}
&= \bigl[\mathbf H_{\mathrm u}^{(1,0)}\bigr]_{AB}
\\
&\quad-\sum_{CD}
\bigl[\mathbf S_{\mathrm u}^{(1,0)}\bigr]_{AC}
\bigl[(\mathbf S_{\mathrm u}^{(0,0)})^{-1}\bigr]_{CD}
\bigl[\mathbf H_{\mathrm u}^{(0,0)}\bigr]_{DB},
\nonumber
\end{align}
\begin{align}
\bigl[\mathbf H^{(0,1)}\bigr]_{AB}
&= \bigl[\mathbf H_{\mathrm u}^{(0,1)}\bigr]_{AB}
\\
&\quad-\sum_{CD}
\bigl[\mathbf H_{\mathrm u}^{(0,0)}\bigr]_{AC}
\bigl[(\mathbf S_{\mathrm u}^{(0,0)})^{-1}\bigr]_{CD}
\bigl[\mathbf S_{\mathrm u}^{(0,1)}\bigr]_{DB},
\nonumber
\end{align}
\begin{align}
\bigl[\mathbf H^{(1,1)}\bigr]_{AB}
&= \bigl[\mathbf H_{\mathrm u}^{(1,1)}\bigr]_{AB}
\\
&\quad-\sum_{CD}
\bigl[\mathbf S_{\mathrm u}^{(1,0)}\bigr]_{AC}
\bigl[(\mathbf S_{\mathrm u}^{(0,0)})^{-1}\bigr]_{CD}
\bigl[\mathbf H_{\mathrm u}^{(0,1)}\bigr]_{DB}
\nonumber\\
&\quad-\sum_{CD}
\bigl[\mathbf H_{\mathrm u}^{(1,0)}\bigr]_{AC}
\bigl[(\mathbf S_{\mathrm u}^{(0,0)})^{-1}\bigr]_{CD}
\bigl[\mathbf S_{\mathrm u}^{(0,1)}\bigr]_{DB}
\nonumber\\
&\quad+\sum_{CDEF}
\bigl[\mathbf S_{\mathrm u}^{(1,0)}\bigr]_{AC}
\bigl[(\mathbf S_{\mathrm u}^{(0,0)})^{-1}\bigr]_{CD}
\bigl[\mathbf H_{\mathrm u}^{(0,0)}\bigr]_{DE}
\nonumber\\
&\qquad\quad\times
\bigl[(\mathbf S_{\mathrm u}^{(0,0)})^{-1}\bigr]_{EF}
\bigl[\mathbf S_{\mathrm u}^{(0,1)}\bigr]_{FB},
\nonumber
\end{align}
and the corresponding overlap terms are
\begin{align}
\bigl[\mathbf S^{(0,0)}\bigr]_{AB}
&= \bigl[\mathbf S_{\mathrm u}^{(0,0)}\bigr]_{AB},
\end{align}
\begin{align}
\bigl[\mathbf S^{(1,0)}\bigr]_{AB}
&= \bigl[\mathbf S^{(0,1)}\bigr]_{AB}=0,
\end{align}

\begin{align}
\bigl[\mathbf S^{(1,1)}\bigr]_{AB}
&= \bigl[\mathbf S_{\mathrm u}^{(1,1)}\bigr]_{AB}
\\
&\quad-\sum_{CD}
\bigl[\mathbf S_{\mathrm u}^{(1,0)}\bigr]_{AC}
\bigl[(\mathbf S_{\mathrm u}^{(0,0)})^{-1}\bigr]_{CD}
\bigl[\mathbf S_{\mathrm u}^{(0,1)}\bigr]_{DB}.
\nonumber
\end{align}

\section{Reduction of NOCI-MP2 to NOCI-RS2}
\label{app:reduction}

In the notation of this work, the unprojected NOCI-MP2 Hamiltonian and overlap matrices of Refs.~\citenum{yostMultireferencePerturbationMethod2013} and \citenum{yostSizeConsistentFormulations2016} take the following forms.
For NOCI-MP2 v0,
\begin{align}
\bigl[\mathbf H_{\mathrm u}^{\mathrm{v0}}\bigr]_{AB}
&= \bigl[\mathbf H_{\mathrm u}^{(0,0)}\bigr]_{AB} \\
&\quad+\frac12\Bigl(
\bigl[\mathbf H_{\mathrm u}^{(0,1)}\bigr]_{AB}
+\bigl[\mathbf H_{\mathrm u}^{(1,0)}\bigr]_{AB}
\Bigr), \nonumber
\end{align}
\begin{align}
\bigl[\mathbf S_{\mathrm u}^{\mathrm{v0}}\bigr]_{AB}
&= \bigl[\mathbf S_{\mathrm u}^{(0,0)}\bigr]_{AB} \\
&\quad+\frac12\Bigl(
\bigl[\mathbf S_{\mathrm u}^{(0,1)}\bigr]_{AB}
+\bigl[\mathbf S_{\mathrm u}^{(1,0)}\bigr]_{AB}
\Bigr). \nonumber
\end{align}
NOCI-MP2 v1 modifies these expressions to give
\begin{align}
\bigl[\mathbf H_{\mathrm u}^{\mathrm{v1}}\bigr]_{AB}
&= \bigl[\mathbf H_{\mathrm u}^{\mathrm{v0}}\bigr]_{AB} \\
&\quad-\frac12\Bigl(
\bigl[\mathbf H_{\mathrm u}^{(0,0)}\bigr]_{AA}
\bigl[\mathbf S_{\mathrm u}^{(0,1)}\bigr]_{AB} \nonumber\\
&\qquad\quad+
\bigl[\mathbf S_{\mathrm u}^{(1,0)}\bigr]_{AB}
\bigl[\mathbf H_{\mathrm u}^{(0,0)}\bigr]_{BB}
\Bigr), \nonumber
\end{align}
\begin{align}
\bigl[\mathbf S_{\mathrm u}^{\mathrm{v1}}\bigr]_{AB}
&= \bigl[\mathbf S_{\mathrm u}^{(0,0)}\bigr]_{AB},
\end{align}
whereas NOCI-MP2 v2 uses
\begin{align}
\bigl[\mathbf H_{\mathrm u}^{\mathrm{v2}}\bigr]_{AB}
&= \bigl[\mathbf H_{\mathrm u}^{\mathrm{v0}}\bigr]_{AB} \\
&\quad+\frac12\Bigl(
\bigl[\mathbf H_{\mathrm u}^{(1,0)}\bigr]_{AA}
\bigl[\mathbf S_{\mathrm u}^{(0,1)}\bigr]_{AB} \nonumber\\
&\qquad\quad+
\bigl[\mathbf S_{\mathrm u}^{(1,0)}\bigr]_{AB}
\bigl[\mathbf H_{\mathrm u}^{(0,1)}\bigr]_{BB}
\Bigr), \nonumber
\end{align}
\begin{align}
\bigl[\mathbf S_{\mathrm u}^{\mathrm{v2}}\bigr]_{AB}
&= \bigl[\mathbf S_{\mathrm u}^{\mathrm{v0}}\bigr]_{AB}.
\end{align}

Projected variants of these methods are obtained by evaluating the same expressions with the projected wavefunction corrections.
This amounts to replacing each unprojected Hamiltonian and overlap coefficient by its projected counterpart,
$\mathbf H_{\mathrm u}^{(i,j)}\rightarrow\mathbf H^{(i,j)}$
and
$\mathbf S_{\mathrm u}^{(i,j)}\rightarrow\mathbf S^{(i,j)}$.

The model-space intermediate normalization enforced by projection gives
$\mathbf S^{(i,0)}=\mathbf S^{(0,i)}=0$ for $i>0$.
Consequently, the terms distinguishing the three NOCI-MP2 variants vanish, and their Hamiltonian and overlap matrices reduce to
\begin{align}
\mathbf H^{\mathrm{v0}}
&=\mathbf H^{\mathrm{v1}}=\mathbf H^{\mathrm{v2}}\\
&=\mathbf H^{(0,0)}
+\frac12\Bigl(\mathbf H^{(0,1)}+\mathbf H^{(1,0)}\Bigr), \nonumber
\end{align}
\begin{align}
\mathbf S^{\mathrm{v0}}
&=\mathbf S^{\mathrm{v1}}=\mathbf S^{\mathrm{v2}}=\mathbf S^{(0,0)}.
\end{align}
Orthogonalizing this common generalized eigenvalue problem with $\mathbf X^{(0,0)}$ yields the NOCI-RS2 Hamiltonian in Eq.~\ref{eq:NOCI_RS2_full}, establishing the equivalence of all three projected NOCI-MP2 variants to NOCI-RS2.

\section*{References}
\bibliography{references}

@string{bj = {Biophys.\ J.}}

@string{cpl = {Chem.\ Phys.\ Lett.}}

@string{cr = {Chem.\ Rev.}}

@string{ijqc = {Int.\ J. Quantum Chem.}}

@string{jcp = {J.~Chem.\ Phys.}}

@string{jctc = {J.~Chem.\ Theory Comput.}}

@string{jpc = {J.~Phys.\ Chem.}}

@string{jpca = {J.~Phys.\ Chem.~A}}

@string{jpcl = {J.~Phys.\ Chem.\ Lett.}}

@string{ledpm = {Lond.\ Edinb.\ Dubl.\ Phil.\ Mag.}}

@string{natcomm = {Nat.\ Commun.}}

@string{pccp = {Phys.\ Chem.\ Chem.\ Phys.}}

@string{prl = {Phys.\ Rev.\ Lett.}}

@string{prog_tp = {Prog.\ Theor.\ Phys.}}

@article{szabadosMulticonfigurationPerturbationTheory2005,
  title = {Multiconfiguration Perturbation Theory: {{Size}} Consistency at Second Order},
  shorttitle = {Multiconfiguration Perturbation Theory},
  author = {Szabados, {\'A}gnes and Rolik, Zolt{\'a}n and T{\'o}th, G{\'a}bor and Surj{\'a}n, P{\'e}ter R.},
  year = 2005,
  month = mar,
  journal = jcp,
  volume = {122},
  number = {11},
  pages = {114104},
  issn = {0021-9606},
  _doi = {10.1063/1.1862235}
}

@article{rolikPerturbationMulticonfigurationWave2003,
  title = {On the Perturbation of Multiconfiguration Wave Functions},
  author = {Rolik, Zolt{\'a}n and Szabados, {\'A}gnes and Surj{\'a}n, P{\'e}ter R.},
  year = 2003,
  month = jul,
  journal = jcp,
  volume = {119},
  number = {4},
  pages = {1922--1928},
  issn = {0021-9606, 1089-7690},
  _doi = {10.1063/1.1584424}
}

@article{leeRegularizedOrbitalOptimizedSecondOrder2018,
  title   = {Regularized Orbital-Optimized Second-Order {M{\o}ller}--{Plesset} Perturbation Theory: {A} Reliable Fifth-Order-Scaling Electron Correlation Model with Orbital Energy Dependent Regularizers},
  author  = {Lee, Joonho and Head-Gordon, Martin},
  journal = jctc,
  volume  = {14},
  number  = {10},
  pages   = {5203--5219},
  year    = {2018},
  _doi     = {10.1021/acs.jctc.8b00731}
}

@article{demoraesNaturalExcitationFramework2026,
  title   = {Natural Excitation Framework for Defining the External Space: Uncontracted and Internally Contracted Multireference Nonorthogonal Wavefunction Theories},
  author  = {de Moraes, Matheus M. F. and Thompson, Lee M.},
  journal = jcp,
  volume  = {164},
  number  = {1},
  pages   = {014107},
  year    = {2026},
  _doi     = {10.1063/5.0304585}
}

@article{lyakhMultireferenceNatureChemistry2012,
  title   = {Multireference Nature of Chemistry: The Coupled-Cluster View},
  author  = {Lyakh, Dmitry I. and Musia{\l}, Monika and Lotrich, Victor F. and Bartlett, Rodney J.},
  journal = cr,
  volume  = {112},
  number  = {1},
  pages   = {182--243},
  year    = {2012},
  _doi     = {10.1021/cr2001417}
}

@article{sundstromNonorthogonalConfigurationInteraction2014,
  title   = {Non-orthogonal configuration interaction for the calculation of multielectron excited states},
  author  = {Sundstrom, Eric J. and Head-Gordon, Martin},
  journal = jcp,
  volume  = {140},
  number  = {11},
  pages   = {114103},
  year    = {2014},
  _doi     = {10.1063/1.4868120}
}

@article{evangelistaCompanionPerturbationTheory2009,
  title   = {A companion perturbation theory for state-specific multireference coupled cluster methods},
  author  = {Evangelista, Francesco A. and Simmonett, Andrew C. and Schaefer, III, Henry F. and Mukherjee, Debashis and Allen, Wesley D.},
  journal = pccp,
  volume  = {11},
  number  = {23},
  pages   = {4728--4741},
  year    = {2009},
  _doi     = {10.1039/b822910d}
}

@article{yostMultireferencePerturbationMethod2013,
  title   = {A multireference perturbation method using non-orthogonal {Hartree}--{Fock} determinants for ground and excited states},
  author  = {Yost, Shane R. and Kowalczyk, Tim and Van Voorhis, Troy},
  journal = jcp,
  volume  = {139},
  number  = {17},
  pages   = {174104},
  year    = {2013},
  _doi     = {10.1063/1.4827456}
}

@article{yostSizeConsistentFormulations2016,
  title   = {Size consistent formulations of the perturb-then-diagonalize {M{\o}ller}--{Plesset} perturbation theory correction to non-orthogonal configuration interaction},
  author  = {Yost, Shane R. and Head-Gordon, Martin},
  journal = jcp,
  volume  = {145},
  number  = {5},
  pages   = {054105},
  year    = {2016},
  _doi     = {10.1063/1.4959794}
}

@article{thomHartreeFockSolutions2009,
  title   = {{Hartree}--{Fock} solutions as a quasidiabatic basis for nonorthogonal configuration interaction},
  author  = {Thom, Alex J. W. and Head-Gordon, Martin},
  journal = jcp,
  volume  = {131},
  number  = {12},
  pages   = {124113},
  year    = {2009},
  _doi     = {10.1063/1.3236841}
}

@article{sanchez-mansillaRoleDynamicElectron2022,
  title   = {On the role of dynamic electron correlation in non-orthogonal configuration interaction with fragments},
  author  = {S{\'a}nchez-Mansilla, A. and Sousa, C. and Kathir, R. K. and Broer, R. and Straatsma, T. P. and de Graaf, C.},
  journal = pccp,
  volume  = {24},
  number  = {19},
  pages   = {11931--11944},
  year    = {2022},
  _doi     = {10.1039/d2cp00772j}
}

@article{lemkeNonorthogonalConfigurationInteraction2025,
  title   = {Nonorthogonal Configuration Interaction of Constraint-Based Orbital-Optimized Excited States: {A} Versatile Method for Theoretical Photochemistry},
  author  = {Lemke, Yannick and Kussmann, J{\"o}rg and Ochsenfeld, Christian},
  journal = jctc,
  volume  = {21},
  number  = {20},
  pages   = {10193--10211},
  year    = {2025},
  _doi     = {10.1021/acs.jctc.5c01064}
}

@article{burtonGeneralizedNonorthogonalMatrix2021,
  title   = {Generalized nonorthogonal matrix elements: Unifying {Wick}'s theorem and the {Slater}--{Condon} rules},
  author  = {Burton, Hugh G. A.},
  journal = jcp,
  volume  = {154},
  number  = {14},
  pages   = {144109},
  year    = {2021},
  _doi     = {10.1063/5.0045442}
}

@article{giulianiPreciseQuantumChemistry2026,
  title   = {Precise Quantum Chemistry calculations with few {Slater} Determinants},
  author  = {Giuliani, Clemens and Nys, Jannes and Martinazzo, Rocco and Carleo, Giuseppe and Rossi, Riccardo},
  journal = natcomm,
  year    = {2026},
  volume = {17},
  eid = {5613},
  _doi     = {10.1038/s41467-026-70255-z}
}

@article{burtonReachingFullCorrelation2020,
  title   = {Reaching Full Correlation through Nonorthogonal Configuration Interaction: {A} Second-Order Perturbative Approach},
  author  = {Burton, Hugh G. A. and Thom, Alex J. W.},
  journal = jctc,
  volume  = {16},
  number  = {9},
  pages   = {5586--5600},
  year    = {2020},
  _doi     = {10.1021/acs.jctc.0c00468}
}

@article{yostEfficientImplementationNOCIMP22018,
  title   = {Efficient Implementation of {NOCI}-{MP2} Using the Resolution of the Identity Approximation with Application to Charged Dimers and Long {C}--{C} Bonds in Ethane Derivatives},
  author  = {Yost, Shane R. and Head-Gordon, Martin},
  journal = jctc,
  volume  = {14},
  number  = {9},
  pages   = {4791--4805},
  year    = {2018},
  _doi     = {10.1021/acs.jctc.8b00697}
}

@article{j.mayhallSpinFlipNonorthogonal2014,
  title   = {Spin--flip non-orthogonal configuration interaction: a variational and almost black-box method for describing strongly correlated molecules},
  author  = {Mayhall, Nicholas J. and Horn, Paul R. and Sundstrom, Eric J. and Head-Gordon, Martin},
  journal = pccp,
  volume  = {16},
  number  = {41},
  pages   = {22694--22705},
  year    = {2014},
  _doi     = {10.1039/C4CP02818J}
}

@article{thomLocatingMultipleSelfConsistent2008,
  title   = {Locating Multiple Self-Consistent Field Solutions: An Approach Inspired by Metadynamics},
  author  = {Thom, Alex J. W. and Head-Gordon, Martin},
  journal = prl,
  volume  = {101},
  number  = {19},
  pages   = {193001},
  year    = {2008},
  _doi     = {10.1103/PhysRevLett.101.193001}
}

@article{hiscockHolomorphicHartreeFock2014a,
  title   = {Holomorphic {Hartree}--{Fock} Theory and Configuration Interaction},
  author  = {Hiscock, Hamish G. and Thom, Alex J. W.},
  journal = jctc,
  volume  = {10},
  number  = {11},
  pages   = {4795--4800},
  year    = {2014},
  _doi     = {10.1021/ct5007696}
}

@article{gilbertSelfConsistentFieldCalculations2008,
  title   = {Self-Consistent Field Calculations of Excited States Using the Maximum Overlap Method ({MOM})},
  author  = {Gilbert, Andrew T. B. and Besley, Nicholas A. and Gill, Peter M. W.},
  journal = jpca,
  volume  = {112},
  number  = {50},
  pages   = {13164--13171},
  year    = {2008},
  _doi     = {10.1021/jp801738f}
}

@article{khaliullinEfficientSelfconsistentField2006,
  title   = {An efficient self-consistent field method for large systems of weakly interacting components},
  author  = {Khaliullin, Rustam Z. and Head-Gordon, Martin and Bell, Alexis T.},
  journal = jcp,
  volume  = {124},
  number  = {20},
  pages   = {204105},
  year    = {2006},
  _doi     = {10.1063/1.2191500}
}

@article{haitExcitedStateOrbital2020,
  title   = {Excited State Orbital Optimization via Minimizing the Square of the Gradient: General Approach and Application to Singly and Doubly Excited States via Density Functional Theory},
  author  = {Hait, Diptarka and Head-Gordon, Martin},
  journal = jctc,
  volume  = {16},
  number  = {3},
  pages   = {1699--1710},
  year    = {2020},
  _doi     = {10.1021/acs.jctc.9b01127}
}

@article{fukutomeTheoryResonatingQuantum1988,
  title   = {Theory of Resonating Quantum Fluctuations in a Fermion System: Resonating {Hartree}-{Fock} Approximation},
  author  = {Fukutome, Hideo},
  journal = prog_tp,
  volume  = {80},
  number  = {3},
  pages   = {417--432},
  year    = {1988},
  _doi     = {10.1143/PTP.80.417}
}

@article{millerNumericallyStableResonating2025,
  title   = {Numerically stable resonating {Hartree}--{Fock}},
  author  = {Miller, Ericka Roy and Parker, Shane M.},
  journal = jcp,
  volume  = {162},
  number  = {10},
  pages   = {104115},
  year    = {2025},
  _doi     = {10.1063/5.0246790}
}

@article{tomitaInitioMolecularOrbital1996,
  title   = {Ab initio molecular orbital calculations by the resonating {Hartree}-{Fock} approach: superposition of non-orthogonal {Slater} determinants},
  author  = {Tomita, Norikazu and Ten-no, Seiichiro and Tanimura, Yoshitaka},
  journal = cpl,
  volume  = {263},
  number  = {5},
  pages   = {687--690},
  year    = {1996},
  _doi     = {10.1016/S0009-2614(96)01266-3}
}

@article{niteLowCostMolecularExcited2019,
  title   = {Low-Cost Molecular Excited States from a State-Averaged Resonating {Hartree}--{Fock} Approach},
  author  = {Nite, Jacob and Jim{\'e}nez-Hoyos, Carlos A.},
  journal = jctc,
  volume  = {15},
  number  = {10},
  pages   = {5343--5351},
  year    = {2019},
  _doi     = {10.1021/acs.jctc.9b00579}
}

@misc{niteEfficientMultiConfigurationalWavefunction,
  title     = {Efficient Multi-Configurational Wavefunction Method with Dynamical Correlation Using Non-Orthogonal Configuration Interaction Singles and Doubles ({NOCISD})},
  author    = {Nite, Jacob and Jim{\'e}nez-Hoyos, Carlos A.},
  howpublished = {ChemRxiv},
  year      = {2019},
  _doi       = {10.26434/chemrxiv.11369646.v1}
}

@article{oosterbaanNonorthogonalConfigurationInteraction2018,
  title   = {Non-orthogonal configuration interaction with single substitutions for the calculation of core-excited states},
  author  = {Oosterbaan, Katherine J. and White, Alec F. and Head-Gordon, Martin},
  journal = jcp,
  volume  = {149},
  number  = {4},
  pages   = {044116},
  year    = {2018},
  _doi     = {10.1063/1.5023051}
}

@article{sunSelectedNonorthogonalConfiguration2024,
  title   = {Selected non-orthogonal configuration interaction with compressed single and double excitations},
  author  = {Sun, Chong and Gao, Fei and Scuseria, Gustavo E.},
  journal = jctc,
  volume  = {20},
  number  = {9},
  pages   = {3741--3748},
  year    = {2024},
  _doi     = {10.1021/acs.jctc.4c00240}
}

@misc{giulianiVariationalLowenergySubspaces2026,
  title         = {Variational low-energy subspaces for chemically accurate excited states},
  author        = {Giuliani, Clemens and Martinazzo, Rocco and Carleo, Giuseppe and Rossi, Riccardo},
  year          = {2026},
  eprint        = {2606.05148},
  archivePrefix = {arXiv},
  primaryClass  = {physics.chem-ph},
  _doi           = {10.48550/arXiv.2606.05148}
}

@article{jafriElectronRepulsionIntegral1974,
  title   = {Electron repulsion integral approximations and error bounds: Molecular applications},
  author  = {Jafri, J. A. and Whitten, J. L.},
  journal = jcp,
  volume  = {61},
  number  = {5},
  pages   = {2116--2121},
  year    = {1974},
  _doi     = {10.1063/1.1682222}
}

@article{whittenCoulombicPotentialEnergy1973,
  title   = {Coulombic potential energy integrals and approximations},
  author  = {Whitten, J. L.},
  journal = jcp,
  volume  = {58},
  number  = {10},
  pages   = {4496--4501},
  year    = {1973},
  _doi     = {10.1063/1.1679012}
}

@article{anderssonSecondorderPerturbationTheory1990,
  title   = {Second-order perturbation theory with a {CASSCF} reference function},
  author  = {Andersson, Kerstin and Malmqvist, Per {\AA}ke and Roos, Bj{\"o}rn O. and Sadlej, Andrzej J. and Wolinski, Krzysztof},
  journal = jpc,
  volume  = {94},
  number  = {14},
  pages   = {5483--5488},
  year    = {1990},
  _doi     = {10.1021/j100377a012}
}

@article{angeliIntroductionNelectronValence2001,
  title   = {Introduction of $n$-electron valence states for multireference perturbation theory},
  author  = {Angeli, C. and Cimiraglia, R. and Evangelisti, S. and Leininger, T. and Malrieu, J.-P.},
  journal = jcp,
  volume  = {114},
  number  = {23},
  pages   = {10252--10264},
  year    = {2001},
  _doi     = {10.1063/1.1361246}
}

@article{angeliNelectronValenceState2002,
  title   = {$n$-electron valence state perturbation theory: {A} spinless formulation and an efficient implementation of the strongly contracted and of the partially contracted variants},
  author  = {Angeli, Celestino and Cimiraglia, Renzo and Malrieu, Jean-Paul},
  journal = jcp,
  volume  = {117},
  number  = {20},
  pages   = {9138--9153},
  year    = {2002},
  _doi     = {10.1063/1.1515317}
}

@article{khedkarModernMultireferenceMethods2021,
  title   = {Modern multireference methods and their application in transition metal chemistry},
  author  = {Khedkar, Abhishek and Roemelt, Michael},
  journal = pccp,
  volume  = {23},
  number  = {32},
  pages   = {17097--17112},
  year    = {2021},
  _doi     = {10.1039/D1CP02640B}
}

@article{siegbahnCompleteActiveSpace1981,
  title   = {The complete active space {SCF} ({CASSCF}) method in a {Newton}--{Raphson} formulation with application to the {HNO} molecule},
  author  = {Siegbahn, Per E. M. and Alml{\"o}f, Jan and Heiberg, Anders and Roos, Bj{\"o}rn O.},
  journal = jcp,
  volume  = {74},
  number  = {4},
  pages   = {2384--2396},
  year    = {1981},
  _doi     = {10.1063/1.441359}
}

@article{malmqvistRestrictedActiveSpace1990,
  title   = {The restricted active space self-consistent-field method, implemented with a split graph unitary group approach},
  author  = {Malmqvist, Per {\AA}ke and Rendell, Alistair and Roos, Bj{\"o}rn O.},
  journal = jpc,
  volume  = {94},
  number  = {14},
  pages   = {5477--5482},
  year    = {1990},
  _doi     = {10.1021/j100377a011}
}

@article{levineCASSCFExtremelyLarge2020,
  title   = {{CASSCF} with Extremely Large Active Spaces Using the Adaptive Sampling Configuration Interaction Method},
  author  = {Levine, Daniel S. and Hait, Diptarka and Tubman, Norm M. and Lehtola, Susi and Whaley, K. Birgitta and Head-Gordon, Martin},
  journal = jctc,
  volume  = {16},
  number  = {4},
  pages   = {2340--2354},
  year    = {2020},
  _doi     = {10.1021/acs.jctc.9b01255}
}

@article{steinAutomatedSelectionActive2016,
  title   = {Automated Selection of Active Orbital Spaces},
  author  = {Stein, Christopher J. and Reiher, Markus},
  journal = jctc,
  volume  = {12},
  number  = {4},
  pages   = {1760--1771},
  year    = {2016},
  _doi     = {10.1021/acs.jctc.6b00156}
}

@article{bensbergCorrespondingActiveOrbital2023,
  title   = {Corresponding Active Orbital Spaces along Chemical Reaction Paths},
  author  = {Bensberg, Moritz and Reiher, Markus},
  journal = jpcl,
  volume  = {14},
  number  = {8},
  pages   = {2112--2118},
  year    = {2023},
  _doi     = {10.1021/acs.jpclett.2c03905}
}

@article{coeInvestigatingMultireferenceCharacter2015,
  title   = {Investigating Multireference Character and Correlation in Quantum Chemistry},
  author  = {Coe, J. P. and Paterson, M. J.},
  journal = jctc,
  volume  = {11},
  number  = {9},
  pages   = {4189--4196},
  year    = {2015},
  _doi     = {10.1021/acs.jctc.5b00543}
}

@article{epifanovskySoftwareFrontiersQuantum2021,
  title   = {Software for the frontiers of quantum chemistry: An overview of developments in the {Q}-{Chem} 5 package},
  author  = {Epifanovsky, Evgeny and Gilbert, Andrew T. B. and Feng, Xintian and Lee, Joonho and Mao, Yuezhi and Mardirossian, Narbe and Pokhilko, Pavel and White, Alec F. and Coons, Marc P. and Dempwolff, Adrian L. and Gan, Zhengting and Hait, Diptarka and Horn, Paul R. and Jacobson, Leif D. and Kaliman, Ilya and Kussmann, J{\"o}rg and Lange, Adrian W. and Lao, Ka Un and Levine, Daniel S. and Liu, Jie and McKenzie, Simon C. and Morrison, Adrian F. and Nanda, Kaushik D. and Plasser, Felix and Rehn, Dirk R. and Vidal, Marta L. and You, Zhi-Qiang and Zhu, Ying and Alam, Bushra and Albrecht, Benjamin J. and Aldossary, Abdulrahman and Alguire, Ethan and Andersen, Josefine H. and Athavale, Vishikh and Barton, Dennis and Begam, Khadiza and Behn, Andrew and Bellonzi, Nicole and Bernard, Yves A. and Berquist, Eric J. and Burton, Hugh G. A. and Carreras, Abel and Carter-Fenk, Kay and Chakraborty, Romit and Chien, Alan D. and Closser, Kristina D. and Cofer-Shabica, Vale and Dasgupta, Saswata and de Wergifosse, Marc and Deng, Jia and Diedenhofen, Michael and Do, Hainam and Ehlert, Sebastian and Fang, Po-Tung and Fatehi, Shervin and Feng, Qingguo and Friedhoff, Triet and Gayvert, James and Ge, Qinghui and Gidofalvi, Gergely and Goldey, Matthew and Gomes, Joe and Gonz{\'a}lez-Espinoza, Cristina E. and Gulania, Sahil and Gunina, Anastasia O. and Hanson-Heine, Magnus W. D. and Harbach, Phillip H. P. and Hauser, Andreas and Herbst, Michael F. and Hern{\'a}ndez Vera, Mario and Hodecker, Manuel and Holden, Zachary C. and Houck, Shannon and Huang, Xunkun and Hui, Kerwin and Huynh, Bang C. and Ivanov, Maxim and J{\'a}sz, {\'A}d{\'a}m and Ji, Hyunjun and Jiang, Hanjie and Kaduk, Benjamin and K{\"a}hler, Sven and Khistyaev, Kirill and Kim, Jaehoon and Kis, Gergely and Klunzinger, Phil and Koczor-Benda, Zsuzsanna and Koh, Joong Hoon and Kosenkov, Dimitri and Koulias, Laura and Kowalczyk, Tim and Krauter, Caroline M. and Kue, Karl and Kunitsa, Alexander and Kus, Thomas and Ladj{\'a}nszki, Istv{\'a}n and Landau, Arie and Lawler, Keith V. and Lefrancois, Daniel and Lehtola, Susi and Li, Run R. and Li, Yi-Pei and Liang, Jiashu and Liebenthal, Marcus and Lin, Hung-Hsuan and Lin, You-Sheng and Liu, Fenglai and Liu, Kuan-Yu and Loipersberger, Matthias and Luenser, Arne and Manjanath, Aaditya and Manohar, Prashant and Mansoor, Erum and Manzer, Sam F. and Mao, Shan-Ping and Marenich, Aleksandr V. and Markovich, Thomas and Mason, Stephen and Maurer, Simon A. and McLaughlin, Peter F. and Menger, Maximilian F. S. J. and Mewes, Jan-Michael and Mewes, Stefanie A. and Morgante, Pierpaolo and Mullinax, J. Wayne and Oosterbaan, Katherine J. and Paran, Garrette and Paul, Alexander C. and Paul, Suranjan K. and Pavo{\v{s}}evi{\'c}, Fabijan and Pei, Zheng and Prager, Stefan and Proynov, Emil I. and R{\'a}k, {\'A}d{\'a}m and Ramos-Cordoba, Eloy and Rana, Bhaskar and Rask, Alan E. and Rettig, Adam and Richard, Ryan M. and Rob, Fazle and Rossomme, Elliot and Scheele, Tarek and Scheurer, Maximilian and Schneider, Matthias and Sergueev, Nickolai and Sharada, Shaama M. and Skomorowski, Wojciech and Small, David W. and Stein, Christopher J. and Su, Yu-Chuan and Sundstrom, Eric J. and Tao, Zhen and Thirman, Jonathan and Tornai, G{\'a}bor J. and Tsuchimochi, Takashi and Tubman, Norm M. and Veccham, Srimukh Prasad and Vydrov, Oleg and Wenzel, Jan and Witte, Jon and Yamada, Atsushi and Yao, Kun and Yeganeh, Sina and Yost, Shane R. and Zech, Alexander and Zhang, Igor Ying and Zhang, Xing and Zhang, Yu and Zuev, Dmitry and Aspuru-Guzik, Al{\'a}n and Bell, Alexis T. and Besley, Nicholas A. and Bravaya, Ksenia B. and Brooks, Bernard R. and Casanova, David and Chai, Jeng-Da and Coriani, Sonia and Cramer, Christopher J. and Cserey, Gy{\"o}rgy and DePrince, III, A. Eugene and DiStasio, Jr., Robert A. and Dreuw, Andreas and Dunietz, Barry D. and Furlani, Thomas R. and Goddard, III, William A. and Hammes-Schiffer, Sharon and Head-Gordon, Teresa and Hehre, Warren J. and Hsu, Chao-Ping and Jagau, Thomas-C. and Jung, Yousung and Klamt, Andreas and Kong, Jing and Lambrecht, Daniel S. and Liang, WanZhen and Mayhall, Nicholas J. and McCurdy, C. William and Neaton, Jeffrey B. and Ochsenfeld, Christian and Parkhill, John A. and Peverati, Roberto and Rassolov, Vitaly A. and Shao, Yihan and Slipchenko, Lyudmila V. and Stauch, Tim and Steele, Ryan P. and Subotnik, Joseph E. and Thom, Alex J. W. and Tkatchenko, Alexandre and Truhlar, Donald G. and Van Voorhis, Troy and Wesolowski, Tomasz A. and Whaley, K. Birgitta and Woodcock, III, H. Lee and Zimmerman, Paul M. and Faraji, Shirin and Gill, Peter M. W. and Head-Gordon, Martin and Herbert, John M. and Krylov, Anna I.},
  journal = jcp,
  volume  = {155},
  number  = {8},
  pages   = {084801},
  year    = {2021},
  _doi     = {10.1063/5.0055522}
}

@article{burtonGeneralizedNonorthogonalMatrix2022,
  title   = {Generalized nonorthogonal matrix elements. {II}: Extension to arbitrary excitations},
  author  = {Burton, Hugh G. A.},
  journal = jcp,
  volume  = {157},
  number  = {20},
  pages   = {204109},
  year    = {2022},
  _doi     = {10.1063/5.0122094}
}

@article{krishnanSelfconsistentMolecularOrbital1980,
  title   = {Self-consistent molecular orbital methods. {XX}. {A} basis set for correlated wave functions},
  author  = {Krishnan, R. and Binkley, J. S. and Seeger, R. and Pople, J. A.},
  journal = jcp,
  volume  = {72},
  number  = {1},
  pages   = {650--654},
  year    = {1980},
  _doi     = {10.1063/1.438955}
}

@article{dunningGaussianBasisSets1989,
  title   = {Gaussian basis sets for use in correlated molecular calculations. {I}. The atoms boron through neon and hydrogen},
  author  = {Dunning, Jr., Thom H.},
  journal = jcp,
  volume  = {90},
  number  = {2},
  pages   = {1007--1023},
  year    = {1989},
  _doi     = {10.1063/1.456153}
}

@article{coulsonXXXIVNotesMolecular1949,
  title   = {{XXXIV}. Notes on the molecular orbital treatment of the hydrogen molecule},
  author  = {Coulson, C. A. and Fischer, I.},
  journal = ledpm,
  volume  = {40},
  number  = {303},
  pages   = {386--393},
  year    = {1949},
  _doi     = {10.1080/14786444908521726}
}

@article{choeIdentifyingRemovingIntruder2001,
  title   = {Identifying and removing intruder states in multireference {M{\o}ller}--{Plesset} perturbation theory},
  author  = {Choe, Yoong-Kee and Witek, Henryk A. and Finley, James P. and Hirao, Kimihiko},
  journal = jcp,
  volume  = {114},
  number  = {9},
  pages   = {3913--3918},
  year    = {2001},
  _doi     = {10.1063/1.1345510}
}

@article{pulayPerspectiveCASPT2Method2011,
  title   = {A perspective on the {CASPT2} method},
  author  = {Pulay, Peter},
  journal = ijqc,
  volume  = {111},
  number  = {13},
  pages   = {3273--3279},
  year    = {2011},
  _doi     = {10.1002/qua.23052}
}

@article{sunPythonSimulationsChemistry2026,
  title   = {The {Python} simulations of chemistry framework: 10 years of an open-source quantum chemistry project},
  author  = {Sun, Qiming and Hermes, Matthew R. and Wu, Xiaojie and Zhai, Huanchen and Zhang, Xing and Ahmed, Abdelrahman M. and Aucar, Juan Jos{\'e} and Backhouse, Oliver J. and Banerjee, Samragni and Bao, Peng and Bogdanov, Nikolay A. and Bystrom, Kyle and Chapoton, Fr{\'e}d{\'e}ric and Chen, Ning-Yuan and Chernyshov, Ivan Yu. and Clifford, Helen S. and Cohen-Janes, Sander and Cui, Zhi-Hao and Damour, Yann D. and Dattani, Nike and Dittmer, Linus Bjarne and Ehlert, Sebastian and Eriksen, Janus Juul and Evangelista, Francesco A. and Ewing, Simon A. and Farahvash, Ardavan and Focke, Kevin and Gao, Yang and Gasperich, Kevin E. and Gillispie, Nathan and Greiner, Jonas and Hennefarth, Matthew R. and Hermann, Jan and Hillenbrand, Christopher and Huhtasalo, Joonatan and Ibrahim, Basil and Jangid, Bhavnesh and Javaremi, Alireza Nejati and Jenkins, Andrew J. and Jin, Yu and King, Daniel S. and Kooi, Derk Pieter and Kurian, Jo S. and Larsson, Henrik R. and Lau, Bryan Tak Gwong and Lee, Seunghoon and Lehtola, Susi and Li, Chenghan and Li, Hao and Li, Jiachen and Li, Rui and Li, Shuhang and Lykhin, Aleksandr O. and Mahajan, Ankit and Mauger, Nastasia and Mazo-Sevillano, Pablo del and Moussa, Jonathan and Nakano, Kousuke and Neufeld, Verena A. and Peng, Linqing and Pham, Hung Q. and Pinski, Peter and Pokhilko, Pavel and Pu, Zhichen and Qian, Yubing and Quiton, Stephen Jon and Schulze, Wanja T. and Scott, Thais R. and Seal, Aniruddha and Serna, James D. and Smith, James E. T. and Smyser, Kori E. and Stahl, Terrence and Sun, Chong and Sung, Kevin J. and Trushin, Egor and Upadhyay, Shiv and Vo, Ethan A. and Vogels, Thijs and Wang, Shirong and Wang, Tai and Wang, Xiao and Wang, Xubo and Wang, Yuanheng and Williamson, Mark and Yang, Junjie and Ye, Hong-Zhou and Yeh, Chia-Nan and Yu, Haiyang and Yu, Jincheng and Yu, Victor Wen-zhe and Zhang, Chaoqun and Zhang, Dayou and Zhang, Yichi and Zhao, Zijun and Zhou, Zehao and Zhu, Andrew J. and Zhu, Tianyu and Berkelbach, Timothy C. and Gagliardi, Laura and Sharma, Sandeep and Sokolov, Alexander Yu. and Chan, Garnet Kin-Lic},
  journal = jcp,
  volume  = {165},
  number  = {10},
  pages   = {102502},
  year    = {2026},
  _doi     = {10.1063/5.0337441}
}

@article{wangMoreNumericalPrecision2025,
  title   = {More Numerical Precision for Less Compute Cost: Optimizing a Local Correlation Algorithm for Second Order {M{\o}ller}--{Plesset} Theory and Comparing against Pair Natural Orbital Methods},
  author  = {Wang, Zhenling and Ling, Haobo and Shi, Tianyi and Shen, Yao and Wang, Zhe and Liu, Yang and Li, Xiaoye S. and Head-Gordon, Martin},
  journal = jctc,
  volume  = {21},
  number  = {21},
  pages   = {10910--10929},
  year    = {2025},
  _doi     = {10.1021/acs.jctc.5c01358}
}

@article{shiLocalSecondOrder2024,
  title   = {Local Second Order {M{\o}ller}-{Plesset} Theory with a Single Threshold Using Orthogonal Virtual Orbitals: {A} Distributed Memory Implementation},
  author  = {Shi, Tianyi and Wang, Zhenling and Aldossary, Abdulrahman and Liu, Yang and Li, Xiaoye S. and Head-Gordon, Martin},
  journal = jctc,
  volume  = {20},
  number  = {18},
  pages   = {8010--8023},
  year    = {2024},
  _doi     = {10.1021/acs.jctc.4c01016}
}

@article{wangSparsityElectronRepulsion2023,
  title   = {Sparsity of the electron repulsion integral tensor using different localized virtual orbital representations in local second-order {M{\o}ller}--{Plesset} theory},
  author  = {Wang, Zhenling and Aldossary, Abdulrahman and Head-Gordon, Martin},
  journal = jcp,
  volume  = {158},
  number  = {6},
  pages   = {064105},
  year    = {2023},
  _doi     = {10.1063/5.0134764}
}

@article{maslenNoniterativeLocalSecond1998,
  title   = {Noniterative local second order {M{\o}ller}--{Plesset} theory: Convergence with local correlation space},
  author  = {Maslen, P. E. and Head-Gordon, M.},
  journal = jcp,
  volume  = {109},
  number  = {17},
  pages   = {7093--7099},
  year    = {1998},
  _doi     = {10.1063/1.477392}
}

@article{wangRegularizedSecondOrderMoller2025a,
  title   = {Regularized Second-Order {M{\o}ller}--{Plesset} Theory: Linear Scaling Implementation and Assessment on Large-Molecule Problems},
  author  = {Wang, Zhenling and Shi, Tianyi and Luo, Weiliang and Kulik, Heather J. and Liu, Yang and Li, Xiaoye S. and Head-Gordon, Martin},
  journal = jctc,
  volume  = {21},
  number  = {14},
  pages   = {6887--6904},
  year    = {2025},
  _doi     = {10.1021/acs.jctc.5c00534}
}

@article{wangLocalSecondOrderMoller2023a,
  title   = {Local Second-Order {M{\o}ller}--{Plesset} Theory with a Single Threshold Using Orthogonal Virtual Orbitals: Theory, Implementation, and Assessment},
  author  = {Wang, Zhenling and Aldossary, Abdulrahman and Shi, Tianyi and Liu, Yang and Li, Xiaoye S. and Head-Gordon, Martin},
  journal = jctc,
  volume  = {19},
  number  = {21},
  pages   = {7577--7591},
  year    = {2023},
  _doi     = {10.1021/acs.jctc.3c00744}
}

@article{pinskiSparseMapsSystematic2015,
  title   = {Sparse maps---{A} systematic infrastructure for reduced-scaling electronic structure methods. {I}. An efficient and simple linear scaling local {MP2} method that uses an intermediate basis of pair natural orbitals},
  author  = {Pinski, Peter and Riplinger, Christoph and Valeev, Edward F. and Neese, Frank},
  journal = jcp,
  volume  = {143},
  number  = {3},
  pages   = {034108},
  year    = {2015},
  _doi     = {10.1063/1.4926879}
}

@article{schutzLoworderScalingLocal1999,
  title   = {Low-order scaling local electron correlation methods. {I}. Linear scaling local {MP2}},
  author  = {Sch{\"u}tz, Martin and Hetzer, Georg and Werner, Hans-Joachim},
  journal = jcp,
  volume  = {111},
  number  = {13},
  pages   = {5691--5705},
  year    = {1999},
  _doi     = {10.1063/1.479957}
}

@article{ayalaLinearScalingSecondorder1999,
  title   = {Linear scaling second-order {M{\o}ller}--{Plesset} theory in the atomic orbital basis for large molecular systems},
  author  = {Ayala, Philippe Y. and Scuseria, Gustavo E.},
  journal = jcp,
  volume  = {110},
  number  = {8},
  pages   = {3660--3671},
  year    = {1999},
  _doi     = {10.1063/1.478256}
}

@article{sheeRegularizedSecondOrderMoller2021,
  title   = {Regularized Second-Order {M{\o}ller}--{Plesset} Theory: {A} More Accurate Alternative to Conventional {MP2} for Noncovalent Interactions and Transition Metal Thermochemistry for the Same Computational Cost},
  author  = {Shee, James and Loipersberger, Matthias and Rettig, Adam and Lee, Joonho and Head-Gordon, Martin},
  journal = jpcl,
  volume  = {12},
  number  = {50},
  pages   = {12084--12097},
  year    = {2021},
  _doi     = {10.1021/acs.jpclett.1c03468}
}

@article{hornUnrestrictedAbsolutelyLocalized2013,
    title = {Unrestricted absolutely localized molecular orbitals for energy decomposition analysis: {Theory} and applications to intermolecular interactions involving radicals},
    volume = {138},
    issn = {0021-9606},
    shorttitle = {Unrestricted absolutely localized molecular orbitals for energy decomposition analysis},
    _url = {https://doi.org/10.1063/1.4798224},
    _doi = {10.1063/1.4798224},
    number = {13},
    urldate = {2026-09-28},
    journal = jcp,
    author = {Horn, Paul R. and Sundstrom, Eric Jon and Baker, Thomas A. and Head-Gordon, Martin},
    month = apr,
    year = {2013},
    pages = {134119},
}
\end{document}